\documentclass[letterpaper,twocolumn,10pt]{article}
\usepackage{usenix}
\usepackage{algorithm,algorithmic}
\usepackage{amsfonts, amsmath, amsthm}
\usepackage[english]{babel}
\usepackage{booktabs}
\usepackage{color}
\usepackage{colortbl}
\usepackage{enumitem}
\usepackage{graphicx}
\usepackage[hidelinks]{hyperref}
\usepackage[utf8]{inputenc}
\usepackage{mathtools}
\usepackage{multirow}
\usepackage{url}
\usepackage[dvipsnames]{xcolor}
\usepackage{xspace}

\DeclarePairedDelimiter{\ceil}{\lceil}{\rceil}

\setlist[itemize]{itemsep=0pt, topsep=0pt, leftmargin=16pt}
\setlist[enumerate]{itemsep=0pt, topsep=0pt, leftmargin=16pt}
\makeatletter
\newcommand{\myitem}[1]{%
\item[#1.]\protected@edef\@currentlabel{#1}%
}
\makeatother

\newcommand{\sys}{\textsf{XRD}\xspace}

\newcommand{\chainsperuser}{\ell}
\newcommand{\msg}{m}
\newcommand{\servers}{N}
\newcommand{\users}{M}

\newcommand{\chain}{k}
\newcommand{\churn}{\gamma}
\newcommand{\chainnum}{\mathsf{x}}
\newcommand{\prfchain}{\mathsf{c}}
\newcommand{\chains}{n}
\newcommand{\inner}{e}

\newcommand{\outerc}{c}
\newcommand{\kex}{\mathsf{DH}}
\newcommand{\public}{\mathsf{pk}}
\newcommand{\private}{\mathsf{sk}}
\newcommand{\secretkey}{\mathsf{s}}
\newcommand{\innerkey}{\mathsf{ipk}}
\newcommand{\privateinnerkey}{\mathsf{isk}}
\newcommand{\blindkey}{\mathsf{bpk}}
\newcommand{\privateblindkey}{\mathsf{bsk}}
\newcommand{\mixingkey}{\mathsf{mpk}}
\newcommand{\privatemixingkey}{\mathsf{msk}}
\newcommand{\round}{\rho}

\newcommand{\nonce}{\mathsf{nonce}}
\newcommand{\aenc}{\mathsf{AEnc}}
\newcommand{\adec}{\mathsf{ADec}}

\newcommand{\know}{\mathsf{KNOW}}
\newcommand{\adv}{\mathsf{A}}

\newcommand{\Z}{\mathbb{Z}}

\begin{document}

\date{}

\title{\sys: Scalable Messaging System with Cryptographic Privacy}

\author{
{\rm Albert Kwon}\\
MIT
\and
{\rm David Lu}\\
MIT PRIMES
\and
{\rm Srinivas Devadas}\\
MIT
} 

\maketitle

\subsection*{Abstract}
Even as end-to-end encrypted communication becomes more popular,
private messaging remains a challenging problem
due to metadata leakages, such as who is communicating with whom.
Most existing systems that hide communication metadata either
(1) do not scale easily, (2) incur significant overheads,
or (3) provide weaker guarantees than cryptographic privacy,
such as differential privacy or heuristic privacy.
This paper presents \sys (short for Crossroads),
a metadata private messaging system
that provides cryptographic privacy,
while scaling easily to support more users by adding more servers.
At a high level, \sys uses multiple mix networks in parallel
with several techniques, 
including a novel technique we call aggregate hybrid shuffle.
As a result, \sys can support 2 million users
with 251 seconds of latency with 100 servers.
This is $12\times$ and $3.7\times$ faster than Atom and Pung, respectively,
which are prior scalable messaging systems with cryptographic privacy.

\section{Introduction} \label{sec:intro}
Many Internet users today have turned to end-to-end encrypted communication
like TLS~\cite{RFC5246} and Signal~\cite{signal}, to protect the \emph{content}
of their communication in the face of widespread surveillance.
While these techniques are starting to see wide adoption,
they unfortunately do not protect the \emph{metadata} of communication,
such as the timing, the size, and the identities of the end-points.
In scenarios where the metadata are sensitive
(e.g., a government officer talking with a journalist for whistleblowing),
encryption alone is not sufficient to protect users' privacy.

Given its importance, there is a rich history of works that
aim to hide the communication metadata,
starting with mix networks (mix-nets)~\cite{mixnet} and
dining-cryptographers networks (DC-Nets)~\cite{dcnet} in the 80s.
Both works provide formal privacy guarantees against global
adversaries, which has inspired many systems
with strong security guarantees~\cite{dissentv1,dissentv2,riffle,vuvuzela}.
However, mix-nets and DC-nets require the users' messages
to be processed by either centralized servers
or every user in the system,
making them difficult to scale to millions of users.
Systems that build on them
typically inherit the scalability limitation as well,
with overheads increasing (often superlinearly)
with the number of users or servers~\cite{dissentv1,dissentv2,riffle,vuvuzela}.
For private communication systems, however, supporting a large user base
is imperative to providing strong security;
as aptly stated by prior works,
``anonymity loves company''~\cite{dingledine2006anonymity,crowds}.
Intuitively, the adversary's goal of learning information about a user
naturally becomes harder as the number of users increases.

As such, many recent messaging systems have been targeting scalability
as well as formal security guarantees.
Systems like Stadium~\cite{stadium} and Karaoke~\cite{karaoke}, for instance, use
differential privacy~\cite{dwork}
to bound the information leakage on the metadata.
Though this has allowed the systems to scale to more users
with better performance,
both systems leak a small bounded amount of metadata for each message,
and thus have a notion of ``privacy budget''.
A user in these systems then spends a small amount of privacy budget
every time she sends a sensitive message,
and eventually is not guaranteed strong privacy.
Users with high volumes of communication could quickly
exhaust this budget, and there is no clear mechanism
to increase the privacy budget once it runs out.
Scalable systems that provide stronger cryptographic privacy
like Atom~\cite{atom} or Pung~\cite{pung},
on the other hand,
do not have such a privacy budget.
However, they rely heavily on expensive cryptographic primitives
such as public key encryption
and private information retrieval~\cite{xpir}.
As a result, they suffer from high latency,
in the order of ten minutes or longer for a few million users,
which impedes their adoption.

This paper presents a point-to-point metadata private messaging
system called \sys 
that aims to marry the best aspects of prior systems.
Similar to several recent works~\cite{pung,atom,stadium},
\sys scales with the number of servers.
At the same time, the system cryptographically hides
all communication metadata from an adversary who
controls the entire network, a constant fraction of the servers,
and any number of users.
Consequently, it can support virtually unlimited amount of communication
without leaking privacy against such an adversary.
Moreover, \sys only uses cryptographic primitives that are significantly
faster than the ones used by prior works,
and can thus provide lower latency and higher throughput
than prior systems with cryptographic security.

In more detail, a \sys deployment consists of many servers.
These servers are organized into many small chains,
each of which acts as a local mix-net.
Before any communication, each user creates a \emph{mailbox}
that is uniquely associated with her,
akin to an e-mail address.
In order for two users Alice and Bob to have a conversation
in the system,
they first pick a number of chains using a specific algorithm
that \sys provides.
The algorithm guarantees that
\emph{every} pair of users intersects at one of the chains.
Then, Alice and Bob send messages addressed to their own mailboxes
to all chosen chains,
except to the chain where their choices of chains align,
where they send their messages for each other.
Once all users submit their messages,
each chain shuffles and decrypts the messages,
and forwards the shuffled messages to the appropriate mailboxes.
Intuitively, \sys protects the communication metadata because
(1) every pair of users meets at a chain
which makes it equally likely for any pair of users to be communicating,
and (2) the mix-net chains hide whether a user sent a message to another user or herself.

One of the main challenges of \sys is addressing active attacks
by malicious servers, where they tamper with some of the users' messages.
This challenge is not new to our system,
and several prior works have employed expensive cryptographic
primitives like verifiable shuffle~\cite{dissentv1,dissentv2,riffle,stadium,atom}
or incurred significant bandwidth overheads~\cite{atom}
to prevent such attacks.
In \sys, we instead propose a new technique called
\emph{aggregate hybrid shuffle}
that can verify the correctness of shuffling
using only efficient cryptographic techniques.

\sys has two significant drawbacks compared to prior systems.
First, with $\servers$ servers,
each user must send $O(\sqrt{\servers})$ messages
in order to ensure that every pair of users intersects.
Second, because each user sends $O(\sqrt{\servers})$ messages,
the workload of each \sys server is $O(\users/\sqrt{\servers})$
for $\users$ users,
rather than $O(\users/\servers)$ like
most prior scalable messaging systems~\cite{pung,atom,stadium}.
Thus, prior systems could outperform \sys
in deployment scenarios with large numbers of servers and users,
since the cost of adding a single user is higher
and adding servers is not as beneficial in \sys.

Nevertheless, our evaluation suggests that
\sys outperforms prior systems with cryptographic guarantees
if there are less than a few thousand servers in the network.
\sys can handle
2 million users (comparable to the number of daily Tor users~\cite{tor_metric})
in 251 seconds with 100 servers.
For Atom~\cite{atom} and Pung~\cite{pung,sealpir},
two prior scalable messaging systems with cryptographic privacy,
it would take over 50 minutes and 15 minutes, respectively.
(These systems, however, can defend against stronger adversaries,
as we detail in \S\ref{sec:related} and \S\ref{sec:eval}.)
Moreover, the performance gap grows with more users,
and we estimate that Atom and Pung require at least
1,000 servers in the network to achieve comparable latency
with 2 million or more users.
While promising, we find that \sys is not as fast
as systems with weaker security guarantees:
Stadium~\cite{stadium} and Karaoke~\cite{karaoke}, for example,
would be $3.3\times$ and $25\times$ faster than \sys, respectively,
in the same deployment scenario.
In terms of user costs,
we estimate that 40~Kbps of bandwidth is sufficient for users
in a \sys network with 2,000 servers,
and the bandwidth requirement scales down to 1~Kbps with 100 servers.

In summary, we make the following contributions:
\begin{itemize}
  \item Design and analyze \sys, a metadata private messaging system
    that can scale by distributing the workload across
    many servers while providing cryptographic privacy.
  \item Design a technique called aggregate hybrid shuffle
    that can efficiently protect users' privacy under active attacks.
  \item Implement and evaluate a prototype of \sys
    on a network of commodity servers,
    and show that \sys outperforms existing cryptographically secure designs.
\end{itemize}

\section{Related work} \label{sec:related}
In this section, we discuss related work
by categorizing the prior systems primarily
by their privacy properties,
and also discuss the scalability and performance of each system.

\paragraph{Systems with cryptographic privacy.}
Mix-nets~\cite{mixnet} and DC-Nets~\cite{dcnet}
are the earliest examples of works that provide
cryptographic (or even information theoretic)
privacy guarantees against global adversaries.
Unfortunately, they have two major issues.
First, they are weak against active attackers:
adversaries can deanonymize users in mix-nets
by tampering with messages,
and can anonymously deny service in DC-Nets.
Second, they do not scale to large numbers of users
because all messages
must be processed by either a small number of servers
or every user in the system.
Many systems that improved on the security of these systems
against active attacks~\cite{dissentv1,dissentv2,riffle,vuvuzela}
suffer from similar scalability bottlenecks.
Riposte~\cite{riposte}, a system that uses ``private information storage''
to provide anonymous broadcast,
also requires all servers to handle a number of messages proportional
to the number of users, and thus faces similar scalability issues.

A recent system Atom~\cite{atom} targets both scalability
and strong anonymity.
Specifically, Atom can scale \emph{horizontally},
allowing it to scale to larger numbers of users simply
by adding more servers to the network.
At the same time, it provides sender anonymity~\cite{pfitzmann2010terminology}
(i.e., no one, including the recipients, learns who sent which message)
against an adversary that can compromise any fraction of the servers
and users.
However, Atom employs expensive cryptography,
and requires the message to be routed
through hundreds of servers in series.
Thus, Atom incurs high latency,
in the order of tens of minutes for a few million users.

Pung~\cite{pung,sealpir} is a system that aims to
provide metadata private messaging between honest users
with cryptographic privacy.
This is a weaker notion of privacy than that of Atom,
as the recipients (who are assumed to be honest)
learn the senders of the messages.
However, unlike most prior works,
Pung can provide private communication even if \emph{all servers}
are malicious by using
a cryptographic primitive called computational
private information retrieval (CPIR)~\cite{chor-pir,xpir}.
Its powerful threat model comes unfortunately at the cost of performance:
Though Pung scales horizontally,
the amount of work required per user is proportional
to the total number of users, resulting in the total work
growing superlinearly with the number of users.
Moreover, PIR is computationally expensive,
resulting in throughput of only a few hundred or
thousand messages per minute per server.

\paragraph{Systems with differential privacy.}
Vuvuzela~\cite{vuvuzela}
and its horizontally scalable siblings Stadium~\cite{stadium}
and Karaoke~\cite{karaoke}
aim to provide differentially private (rather than
cryptographically private) messaging.
At a high level, they hide the communication patterns of honest users
by inserting dummy messages that are indistinguishable
from real messages, and reason carefully about
how much information is leaked at each round.
They then set the system parameters such that they could support
a number of sensitive messages; for instance, Stadium and Karaoke
target $10^4$ and $10^6$ messages, respectively.
Up to that number of messages, the systems
allow users to provide a plausible cover story
to ``deny'' their actual actions.
Specifically, the system ensures that
the probability of Alice conversing with Bob from the adversary's perspective
is within $e^\epsilon$
(typically, $e^\epsilon \in [3,10]$)
of the probability of Alice conversing with any other user
with only a small failure probability $\delta$
(typically, $\delta=0.0001$).
This paradigm shift has allowed
the systems to support larger numbers of users
with lower latency than prior works.

Unfortunately, systems with differential privacy suffer from two drawbacks.
First, the probability gap between two events
may be sufficient for strong adversaries to act on.
For instance, if Alice is ten times as likely to talk to Bob
than Charlie, the adversary may act assuming that
Alice is talking to Bob, despite the plausible deniability.
Second, there is a ``privacy budget'' (e.g., $10^4$ to $10^6$ messages),
meaning that a user can deny a limited number of messages
with strong guarantees.
Moreover, for best possible security, users must constantly
send messages, and deny every message.
For instance, Alice may admit that she is not in any conversation
(thinking this information is not sensitive),
but this could have unintended consequences
on the privacy of another user
who uses the cover story that she is talking with Alice.
The budget could then run out quickly if users want the strongest privacy possible:
If a user sends a message every minute,
she would run out of her budget in a few days or years
with $10^4$ to $10^6$ messages.
Although the privacy guarantee weakens gradually
after the privacy budget is exhausted,
it is unclear how to raise the privacy levels once they have been lowered.

\paragraph{Scalable systems with other privacy guarantees.}
The only private communication system in wide-deployment today
is Tor~\cite{tor}. Tor currently supports over
2 million daily users using over 6,000 servers~\cite{tor_metric},
and can scale to more users easily by adding more servers.
However, Tor does not provide privacy against an adversary
that monitors significant portions of the network, and is susceptible
to traffic analysis attacks~\cite{tor_traffic_analysis,ccs2013-usersrouted}.
Its privacy guarantee weakens further
if the adversary can control some servers,
and if the adversary launches active attacks~\cite{correlated_flow}.
Similar to Tor, most free-route mix-nets~\cite{shadowwalker,tarzan,crowds,mixminion,aqua}
(distributed mix-nets where each messages is routed through
a small subset of servers)
cannot provide strong privacy against powerful adversaries
due to traffic analysis and active attacks.

Loopix~\cite{loopix} is a recent iteration on free-route mix-nets,
and can provide fast asynchronous messaging.
To do so, each user interacts with a semi-trusted server
(called ``provider'' in the paper),
and routes her messages through a small number of servers
(e.g., 3 servers).
Each server inserts small amounts of random delays before routing
the messages.
Loopix then reasons about privacy using entropy.
Unfortunately, the privacy guarantee of Loopix weakens
quickly as the adversary compromises more servers.
Moreover, Loopix requires the recipients to trust the provider
to protect themselves.

\section{System model and goals} \label{sec:model}
\sys aims to achieve the best of all worlds by
providing cryptographic metadata privacy
while scaling horizontally
without relying on expensive cryptographic primitives.
In this section, we present our threat model
and system goals.

\subsection{Threat model and assumptions} \label{sec:threat_model}
A deployment of \sys would consist of hundreds to thousands
of servers and a large number of users, in the order of millions.
Similar to several prior works on distributed
private communication systems~\cite{atom,stadium},
\sys assumes an adversary that can monitor the entire network,
control a fraction $f$ of the servers,
and control up to all but two honest users.
We assume, however, that there exists
a public key infrastructure that can be used to
securely share public keys of online servers and users
with all participants at any given time.
These keys, for example, could be maintained by key transparency
schemes~\cite{RFC6962,coniks,catena}.

\sys does not hide the fact that users are using \sys.
Thus, for best possible security, users should stay online to avoid
intersection attacks~\cite{kedogan_intersection,danezis_intersection}.
\sys also does not protect against
denial-of-service (DoS) attacks,
and in general does not provide strong availability guarantees.
Defending against intersection and DoS attacks is an interesting future work.
\sys does, however,
provide privacy even under DoS, server churn,
and user churn
(e.g., Alice goes offline unexpectedly
without her conversation partner knowing).
We discuss the availability properties further
in \S\ref{sec:design} and \S\ref{sec:availability}.

Finally, \sys assumes that the users can agree to start
talking at a certain time out-of-band.
This could be done, for example, via two users
exchanging this information offline,
or by using systems like Alpenhorn~\cite{alpenhorn}
that can initiate conversations privately.

\paragraph{Cryptographic primitives.} 
\sys uses standard cryptographic primitives.
It assumes existence of a group of prime order $p$ with
a generator $g$ in which discrete log is hard and
the decisional Diffie-Hellman assumption holds.
We will write $\kex(g^a, b) = g^{ab}$
to denote Diffie-Hellman key exchange.
In addition, \sys makes use of authenticated encryption.

\textbf{Authenticated encryption~\cite{cryptoeprint:2003:069}:}
\sys relies on an authenticated encryption scheme for confidentiality
and integrity, which consists of the following algorithms:
\begin{itemize}
  \item $\outerc \leftarrow \aenc(\secretkey, \nonce, \msg)$.
    Encrypt message $\msg$ and authenticate the ciphertext $\outerc$
    using a symmetric key $\secretkey$ and a nonce $\nonce$.
    Typically, $\secretkey$ is used to derive two more keys,
    one for encryption and one for authentication (e.g., via HMAC).
  \item $(b, \msg) \leftarrow \adec(\secretkey, \nonce, \outerc)$.
    Check the integrity of and decrypt ciphertext $\outerc$ using
    the key $\secretkey$ and a nonce $\nonce$.
    If the check fails, then $b=0$ and $\msg=\bot$.
    Otherwise, $b=1$ and $\msg$ is the underlying plaintext.
\end{itemize}
\sys in particular relies on the following events having negligible
probability when using authenticated encryption:
(1) generating a correctly authenticated ciphertext without knowing
the secret key used for $\adec$,
and (2) the same ciphertext authenticating under two different keys.
Both of these properties are true, for example,
when using the encrypt-then-MAC authenticated encryption schemes.

\subsection{Goals} \label{sec:goals}

\sys has three main goals.

\paragraph{Correctness.}
Informally, the system is correct if
every honest user successfully communicates with
her conversation partner
after a successful execution of the system protocol.

\paragraph{Privacy.}
Similar to prior messaging systems~\cite{vuvuzela,pung,stadium,karaoke},
\sys aims to provide relationship unobservability~\cite{pfitzmann2010terminology},
meaning that the adversary cannot learn anything
about the communication between two honest users.
Informally, consider any honest users Alice, Bob, and Charlie.
The system provides privacy
if the adversary cannot distinguish whether Alice is communicating
with Bob, Charlie, or neither.
\sys only guarantees this property among the honest users,
as malicious conversation partners
can trivially learn the metadata of their communication.
We provide a more formal definition in Appendix~\ref{sec:security}.
(This is a weaker privacy goal than that of Atom~\cite{atom},
which aims for sender anonymity.)

\paragraph{Scalability.}
Similar to prior work~\cite{atom},
we require that the system can handle more users with more servers.
If the number of messages processed by a server
is $C(\users, \servers)$ for $\users$ users and $\servers$ servers,
we require that
$C(\users, \servers) \rightarrow 0$ as $\servers \rightarrow \infty$.
$C(\users, \servers)$ should approach zero
polynomially in $\servers$ so that adding a server introduces
significant performance benefits.

\section{\sys overview} \label{sec:overview}
\begin{figure}[t]
  \centering
  \includegraphics[width=\linewidth]{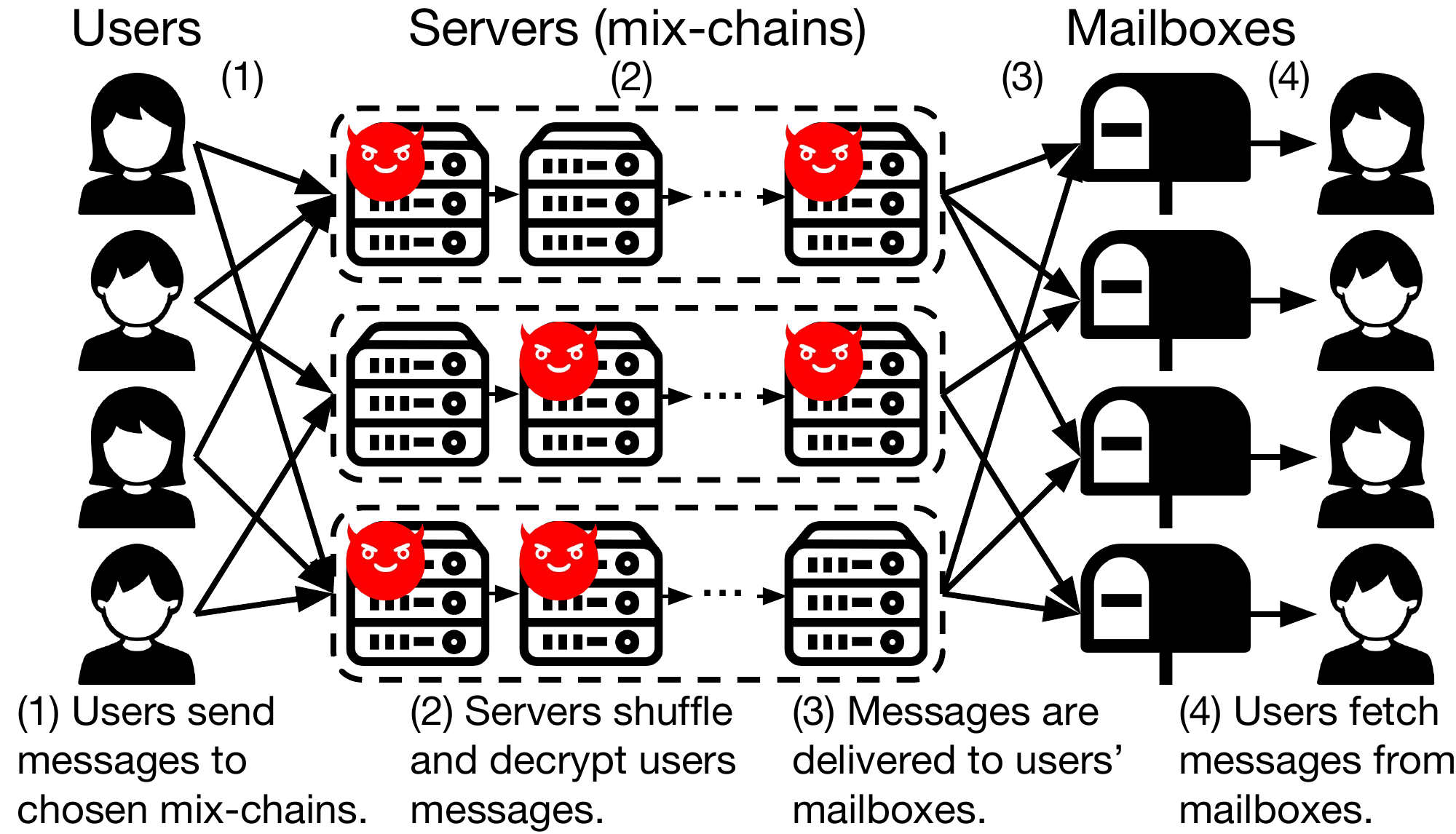}
  \caption{Overview of \sys operation.}
  \label{fig:overview}
\end{figure}

Figure~\ref{fig:overview} presents the overview of a \sys network.
At a high level, \sys consists of three different entities:
users, mix servers, and mailbox servers.
Every user in \sys has a unique \emph{mailbox} associated
with her, similar to an e-mail address.
The mailbox servers maintain the mailboxes,
and are only trusted for availability and not privacy.

To set up the network,
\sys organizes the mix servers into many chains of
servers such that there exists at least one honest server
in each chain with overwhelming probability
(i.e., an anytrust group~\cite{dissentv2}).
Communication in \sys is carried out in discrete rounds.
In each round, each user selects a fixed set of $\chainsperuser$ chains,
where the set is determined by the user's public key.
She then sends a fixed size message to each of the selected chains.
(If the message is too small or large,
then the user pads the message or breaks it into multiple pieces.)
Each message contains a destination mailbox,
and is onion-encrypted for all servers in the chain.

Once all users' messages are submitted,
each chain acts as a local mix-net~\cite{mixnet},
decrypting and shuffling messages.
During shuffling, each server also generates a short proof
that allows other servers to check that it behaved correctly.
If the proof does not verify,
then the protocol halts with no privacy leakage.
If all verification succeeds,
then the last server in each chain forwards
the messages to the appropriate mailbox servers.
(The protocol is described in detail in \S\ref{sec:active}.)
Finally, the mailbox servers put the messages into the appropriate mailboxes,
and each user downloads
all messages in her mailbox at the end of a round.

The correctness and security of \sys is in large part due
to how each user selects the mix-nets and the destination
of each message.
As we will see in \S\ref{sec:design}, the users are required
to follow a specific algorithm to select the chains.
The algorithm guarantees that \emph{every} pair of users
have at least one chain in common
and the choices of the chains are publicly computable.
For example, every user selecting the same chain will achieve this property,
and thus correctness and security.
In \sys, we achieve this property while distributing the load evenly.

Let us now consider two scenarios:
(1) a user Alice is not in a conversation with anyone,
or (2) Alice is in a conversation with another user Bob.
In the first case,
she sends a dummy message encrypted for herself to each chain
that will come back to her own mailbox.
We call these messages \emph{loopback messages}.
In the second case,
Alice and Bob compute each other's choices of chains,
and discover at which chain they will intersect.
If there are multiple such chains,
they break ties in a deterministic fashion.
Then, Alice and Bob send the messages
encrypted for the other person,
which we call \emph{conversation messages},
to their intersecting chain.
They also send loopback messages on all other chains.

\subsection{Security properties} \label{sec:security_overview}
We now argue the security informally.
We present a more formal definition and arguments
of privacy in Appendix~\ref{sec:security}.
Since both types of messages are encrypted for owners of mailboxes
and the mix-net hides the origin of a message,
the adversary cannot tell if a message going to Alice's mailbox
is a loopback message or a conversation message sent by a different user.
This means that the network pattern of all users is the same
from the adversary's perspective: each user sends and receives
exactly $\chainsperuser$ messages, each of which could be a loopback
or a conversation message.
As a result, the adversary cannot tell if a user is in a conversation or not.
Moreover, we choose the chains such that
every pair of users intersects at some chain (\S\ref{sec:design}),
meaning the probability that Alice is talking to another honest
user is the same for all honest users.
This hides the conversation metadata.

The analysis above, however, only holds if the adversary does
not tamper with the messages. For instance, if the
adversary drops Alice's message in a chain,
then there are two possible observable outcomes in this chain:
Alice receives (1) no message,
meaning Alice is not in a conversation in this chain, or
(2) one message, meaning someone intersecting with Alice
at this chain is chatting with Alice.
This information leakage breaks the security of \sys.
We propose a new protocol called aggregate hybrid shuffle (\S\ref{sec:active})
that efficiently defends against such an attack.

\subsection{Scalability properties} \label{sec:scalability}
Let $\chains$ and $\servers$ be the number of chains and servers
in the network, respectively.
Each user must send at least $\sqrt{\chains}$ messages to guarantee every pair
of users intersect. To see why, fix $\chainsperuser$,
the number of chains a user selects.
Those chains must connect a user Alice
to all $\users$ users.
Since the total number of messages sent by users is $\users \cdot \chainsperuser$,
each chain should handle $\frac{\users \cdot \chainsperuser}{\chains}$ messages
if we distribute the load evenly.
We then need
${\frac{\users \cdot \chainsperuser}{\chains} \cdot \chainsperuser \geq \users}$
because the left hand side is the maximum number of users connected to the chains
that Alice chose.
Thus, $\chainsperuser \geq \sqrt{\chains}$.
In \S\ref{sec:design}, we present an approximation algorithm
that uses $\chainsperuser \approx \sqrt{2\chains}$
to ensure all users intersect with each other
while evenly distributing the work.
This means that each chain handles
$\approx \frac{\sqrt{2}\users}{\sqrt{\chains}}$ messages,
and thus \sys scales with the number of chains.
If we set ${\chains = \servers}$
and each server appears in $\chain$ chains for $\chain << \sqrt{\servers}$,
this satisfies our scalability goal in \S\ref{sec:goals}:
${C(\users, \servers) = \frac{\chain\sqrt{2}\users}{\sqrt{\servers}} \rightarrow 0}$
polynomially as ${\servers \rightarrow \infty}$.
We show that $\chain$ is logarithmic in $\servers$
in \S\ref{sec:anytrust}.

\section{\sys design} \label{sec:design}
In this section, we present the detailed operations
of a base \sys design that protects against an adversary
that does \emph{not} launch active attacks.
We then describe modifications to this baseline design
that allows \sys to protect against active attacks
in \S\ref{sec:active}.

\subsection{Mailboxes and mailbox servers} \label{sec:mailbox}
Every user in \sys has a mailbox that is publicly associated with her.
In our design, we use the public key of each user as the
identifier for the mailbox,
though different public identifiers like e-mail addresses
can work as well.
These mailboxes are maintained by the mailbox servers,
with simple put and get functionalities
to add and fetch messages to a mailbox.
Similar to e-mail servers, different users' mailboxes
can be maintained by different servers.

\subsection{Mix chains} \label{sec:mixnet}
\sys uses many parallel mix-nets to process the messages.
We now describe their formation and operations.

\subsubsection{Forming mix chains} \label{sec:anytrust}
We require the existence of an honest server in every chain
to guarantee privacy.
To ensure this property,
we use public randomness sources~\cite{bonneau2015bitcoin, sytascalable}
that are unbiased and publicly available
to randomly sample $\chain$ servers to form a chain,
similar to prior works~\cite{atom,stadium}.
We set $\chain$ large enough such that
the probability that all servers are malicious is negligible.
Concretely, the probability that a chain of length $\chain$
consists only of malicious servers is $f^\chain$.
Then, if we have $\chains$ chains in total,
the probability there exists a group of only malicious servers
is less than $\chains \cdot f^\chain$ via a union bound.
Finally, we can upper bound this to be negligible.
For example, if we want this probability to be less than $2^{-64}$
for $f = 20\%$, then we need $\chain = 32$ for $\chains < 6000$.
This makes $\chain$ depend logarithmically on $\servers$.
In \sys, we set $\chains = \servers$ for $\servers$ servers,
meaning each server appears in $\chain$ chains on average.

Once the servers are selected, we ``stagger'' the position of a server in the chains
to ensure maximal server utilization.
For instance, if a server is part of two chains,
then it could be the first server in one chain and
the second server in the other chain.
This optimization has no impact on the security,
as we only require the existence of an honest server in each group.
This helps minimize the idle time of each server.

\subsubsection{Processing user messages} \label{sec:mix}
\begin{algorithm}[t]
  \caption{Mix server routing protocol} \label{alg:mix}
  Server $i$ in a chain of $\chain$ servers
  possesses its mixing key pair
  $(\mixingkey_i = g^{\privatemixingkey_i}, \privatemixingkey_i)$.
  In each round $\round$, it receives a set of ciphertexts
  ${\{\outerc_i^j = (g^{x_j}, \aenc(\kex(\mixingkey_i, x_j), \round, \outerc_{i+1}^j)\}}$
  for each user $j$,
  either from an upstream server if $i \neq 1$, or from the users if $i = 1$.
  \begin{enumerate}
    \myitem{1} \textbf{Decrypt and shuffle:} \label{step:dec_shuf}
      Decrypt each message:
      ${\outerc_{i+1}^j =
        \adec(\kex(g^{x_j},\privatemixingkey_i), \round, \outerc_i^j)}$.
      Randomly shuffle $\{\outerc_{i+1}^j\}$.
    \myitem{2a} \textbf{Relay messages:}
      If $i < \chain$, then send the shuffled $\{\outerc_{i+1}^j\}$ to server $i+1$.
    \myitem{2b} \textbf{Forward messages to mailbox:} \label{step:forward}
      If $i = \chain$, then each decrypted message is of the form
      $(\public_u, \aenc(\secretkey, \round, \msg_u))$,
      where $\public_u$ is the public key of a user $u$,
      $\secretkey$ is a secret key, and $\msg_u$ is a message for the user.
      Send the message to the mailbox server that manages
      mailbox $\public_u$.
  \end{enumerate}
\end{algorithm}

After the chains are created, each mix server $i$ generates
a \emph{mixing key pair} $(\mixingkey_i = g^{\privatemixingkey_i}, \privatemixingkey_i)$,
where $\privatemixingkey_i$ is a random value in $\Z_p$
and $g$ is a generator of the group.
The public mixing keys $\{\mixingkey_i\}$ are made available to all participants
in the network, along with the ordering of the keys in each chain.
Now, each chain behaves as a simple mix-net~\cite{mixnet}:
users submit some messages onion-encrypted using the mixing keys
(\S\ref{sec:client}), and the servers go in order
decrypting and shuffling the messages.
Algorithm~\ref{alg:mix} describes the protocol in detail.
The honest server in each chain is responsible for
hiding the origins of the messages against passive adversaries.
We then protect against active attacks using a new technique
described in \S\ref{sec:active}.

\subsubsection{Server churn} \label{sec:server_churn}
Some servers may go offline in the middle of a round.
Though \sys does not provide additional fault tolerance mechanisms,
only the chains that contain failing servers are affected.
Furthermore, the failing chains do not affect the security
since they do not disturb the operations of other chains
and the destination of the messages
at the failing chain remains hidden to the adversary.
Thus, conversations that use chains
with no failing servers are unaffected.
We analyze the empirical effects of server failures in \S\ref{sec:availability}.

\subsection{Users} \label{sec:client}
We now describe how users operate in \sys.

\subsubsection{Selecting chains} \label{sec:chain_selection}
\sys needs to ensure that all users' choices of chains
intersect at least once, and that the choices are publicly computable.
We present a scheme that achieves this property.
Upon joining the network, every user is placed into one of
$\chainsperuser+1$ groups such that each group contains roughly
the same number of users,
and such that the group of any user is publicly computable.
This could be done, for example,
by assigning each user to a pseudo-random group
based on the hash of the user's public key.
Every user in a group is connected to the same $\chainsperuser$ servers
specified as follows.
Let $C_i$ be the ordered set of chains that users in group $i$ are connected to.
We start with $C_1 = \{1, \ldots, \chainsperuser\}$,
and build the other sets inductively:
For $i = 1, \ldots, \chainsperuser$,
group $i+1$ is connected to
$C_{i+1} = {\{C_1[i], C_2[i], \ldots, C_{i}[i],
  C_{i}[\chainsperuser] + 1,\ldots, C_{i}[\chainsperuser] + (\chainsperuser - i) \}}$,
where $C_x[y]$ is the $y^{\text{th}}$ entry in $C_x$.

By construction, every group is connected
to every other group: Group $i$ is connected to group $j$
via $C_i[j]$ for all $i < j$. As a result, every user in group $i$
is connected to all others in the same group
(they meet at all chains in $C_i$),
and is connected to users in group $j$ via chain $C_i[j]$.

To find the concrete value of $\chainsperuser$,
let us consider $C_{\chainsperuser}$.
The last chain of $C_{\chainsperuser}$,
which is the chain with the largest index, is
${C_{\chainsperuser}[\chainsperuser] =
  \chainsperuser^2 - \sum_{i=1}^{\chainsperuser-1} i
  = \frac{\chainsperuser^2 + \chainsperuser}{2}}$.
This value should be as close as possible to $\chains$, the number of chains,
to maximize utilization.
Thus, ${\chainsperuser = \ceil{\sqrt{2n+0.25}-0.5}} \approx \ceil{\sqrt{2n}}$.
Given that $\chainsperuser \geq \sqrt{\chains}$ (\S\ref{sec:scalability}),
this is a $\sqrt{2}$-approximation.

\subsubsection{Sending messages} \label{sec:user_conversation}

\begin{algorithm}[t]
  \caption{User conversation protocol} \label{alg:client}
  Consider two users Alice and Bob
  with key pairs ${(\public_A = g^{\private_A}, \private_A)}$
  and ${(\public_B = g^{\private_B}, \private_B)}$
  who are connected to sets of $\chainsperuser$ chains $C_A$ and $C_B$
  (\S\ref{sec:chain_selection}).
  The network consists of chains $1, \ldots, \chains$,
  each with $\chain$ servers.
  Alice and Bob possess
  the set of mixing keys for each chain.
  Alice performs the following in round $\round$.
  \begin{enumerate}
    \myitem{1a} \textbf{Generate loopback messages:}
      If Alice is not in a conversation,
      then Alice generates $\chainsperuser$ loopback messages:
      $\msg_\chainnum = (\public_A, \aenc(\secretkey_A^\chainnum, \round, 0))$
      for $\chainnum \in C_A$,
      where $\secretkey_A^\chainnum$ is
      a chain-specific symmetric key known only to Alice.
    \myitem{1b} \textbf{Generate conversation message:}
      If Alice is in a conversation with Bob,
      then she first computes the shared key
      $\secretkey_{AB} = \kex(\public_B, \private_A)$,
      and the symmetric encryption key for Bob
      $\secretkey_B = \mathsf{KDF}(\secretkey_{AB}, \public_B)$ where
      $\mathsf{KDF}$ is a key derivation function.
      Alice then generates the conversation message:
      ${\msg_{\chainnum_{AB}} = (\public_B, \aenc(\secretkey_{B}, \round, \mathsf{msg}))}$,
      where $\mathsf{msg}$ is the plaintext message for Bob
      and $\chainnum_{AB} \in C_A \cap C_B$ is the first chain
      in the intersection.
      She also generates $\chainsperuser-1$ loopback messages
      $\msg_\chainnum$ for ${\chainnum \in C_A, \chainnum \neq \chainnum_{AB}}$.
    \myitem{2} \textbf{Onion-encrypt messages:} \label{step:onion_enc}
      For each message $\msg_\chainnum$,
      let $\outerc_{\chain+1} = \msg_\chainnum$, and
      let $\{\mixingkey_i\}$ be the mixing keys for chain $\chainnum \in C_A$.
      For $i = \chain$ to 1, generate a random value $x_i \in \Z_p$, and compute
      ${\outerc_i = (g^{x_i}, \aenc(\kex(\mixingkey_i, x_i), \round, \outerc_{i+1}))}$.
      Send $\outerc_1$ to chain $\chainnum$.
    \myitem{3} \textbf{Fetch messages:}
      At the end of the round, fetch and decrypt the messages in her mailbox,
      using $\adec$ with matching $\secretkey_A^\chainnum$ or
      $\secretkey_{A} = \mathsf{KDF}(\secretkey_{AB}, \public_A)$.
  \end{enumerate}
\end{algorithm}

After choosing the $\chainsperuser$ mix chains,
the users send one message to each of the chosen chains
as described in Algorithm~\ref{alg:client}.
At a high level, if Alice is not talking with anyone,
Alice generates $\chainsperuser$ loopback messages
by encrypting dummy messages (e.g., messages with all zeroes)
using a secret key known only to her,
and submits them to the chosen chains.
If she is talking with another user Bob,
then she first finds where they intersect
by computing the intersection of Bob's group
and her group (\S\ref{sec:chain_selection}).
If there is more than one such chain,
then she breaks the tie by selecting the chain with the smallest index.
Alice then generates $\chainsperuser-1$ loopback messages
and one encrypted message using a secret key 
that Alice and Bob shares.
Finally, Alice sends the message for Bob to the intersecting chain,
and sends the loopback messages to the other chains.
Bob mirrors Alice's actions.

\subsubsection{User churn} \label{sec:user_churn}
Like servers, users might go offline in the middle of a round,
and \sys aims to provide privacy in such situations.
However, the protocol presented thus far does not achieve this goal.
If Alice and Bob are conversing
and Alice goes offline without Bob knowing,
then Alice's mailbox will receive Bob's message while Bob's mailbox
will get one fewer message. Thus, by observing mailbox access counts
and Alice's availability, the adversary can infer their communication.

To solve this issue, we require Alice to submit two sets of
messages in round $\round$: the messages for the current round $\round$,
and \emph{cover messages} for round $\round+1$.
If Alice is not communicating with anyone,
then the cover messages will be loopback messages.
If Alice is communicating with another user,
then one of the cover messages will be a conversation message
indicating that Alice has gone offline.

If Alice goes offline in round $\tau$,
then the servers use the cover messages submitted in $\tau-1$
to carry out round $\tau$.
Now, there are two possibilities. If Alice is not in a conversation,
then Alice's cover loopback messages are routed in round $\tau$,
and nothing needs to happen afterwards.
If Alice is conversing with Bob,
then at the end of round $\tau$, Bob will get the message
that Alice is offline via one of the cover messages.
Starting from round $\tau+1$, Bob now sends loopback messages
instead of conversation messages to hide the fact that
Bob was talking with Alice in previous rounds.
This could be used to end conversations as well.
Malicious servers cannot
fool Bob into thinking Alice has gone offline
by replacing Alice's messages with her cover messages
because the honest servers will ensure Alice's real messages
are accounted for using our defenses described in \S\ref{sec:active}.

\section{Aggregate hybrid shuffle} \label{sec:active}

Adversarial servers can tamper with the messages to leak privacy in \sys.
For example, consider a mix-net chain where the first
server is malicious. This malicious server can replace Alice's message
with a message directed at Alice.
Then, at the end of the mixing, the adversary
will make one of two observations.
If Alice was talking to another user Bob,
Bob will receive one fewer message while Alice would receive two messages.
The adversary would then learn that Alice was talking to Bob.
If Alice is not talking to anyone on the tampered chain,
then Alice would receive one message, revealing the lack of conversation
on that chain.

Prior works~\cite{dissentv1,dissentv2,riffle,stadium,atom} have used
traditional verifiable shuffles~\cite{neff,crypto01-shuffle,kdd04-shuffle,groth}
to prevent these attacks.
At a high level, verifiable shuffles allow the servers
in the chain (one of which is honest)
to verify the correctness of a shuffle of another server;
namely, that the plaintexts underlying the outputs of a server
is a valid permutation of the plaintexts underlying the inputs.
Unfortunately, these techniques are computationally expensive,
requiring many exponentiations.

In \sys, we make an observation that help us avoid
traditional verifiable shuffles.
For a meaningful tampering, the adversary necessarily
has to tamper with the messages \emph{before} they are shuffled by
the honest server.
Otherwise, the adversary does not learn the origins of messages.
For example, after dropping a message in a server downstream
from the honest server,
the adversary might observe that Alice did not receive a message.
The adversary cannot tell, however, whether the dropped message
was sent by Alice or another user,
and does not learn anything about Alice's communication pattern.
(Intuitively, the adversarial downstream servers do not add any privacy
in any case.)
In this section, we describe a new form of verifiable shuffle we
call \emph{aggregate hybrid shuffle} (AHS) that allows us to take advantage
of this fact. In particular, the protocol guarantees that the honest
server will receive and shuffle all honest users' messages,
or the honest server will detect that some parties upstream
(some servers or users) misbehaved.
We will then describe how the honest server can efficiently identify
all malicious users and servers who deviated from the protocol,
without affecting the privacy of honest users.
The user and server protocols remain largely the same as the baseline protocols
described in \S\ref{sec:design}, with some crucial changes.

\subsection{Key generation with AHS} \label{sec:active_keys}
When the chain is created, the servers generate three key pairs:
\emph{blinding key}, \emph{mixing key}, and \emph{inner key}.
The inner keys are per-round keys, and
each server $i$ generates its own inner key pair
$(\innerkey_i = g^{\privateinnerkey_i}, \privateinnerkey_i)$.
The other two keys are long-term keys,
and are generated in order starting with the first server in the chain.
Let $\blindkey_0 = g$. Starting with server 1,
server $i = 1, \ldots, \chain$ generates
$(\blindkey_i = \blindkey_{i-1}^{\privateblindkey_i}, \privateblindkey_i)$
and
$(\mixingkey_i = \blindkey_{i-1}^{\privatemixingkey_i}, \privatemixingkey_i)$
in order.
In other words, the base of the public keys of the server $i$
is $\blindkey_{i-1} = g^{\prod_{a<i} \privateblindkey_a}$.
The public mixing key of the last server, for example, would be
$\mixingkey_\chain = \blindkey_{\chain-1}^{\privatemixingkey_\chain}
= g^{\privatemixingkey_\chain \cdot \prod_{i<k} \privateblindkey_i}$.
Each server also has to prove to all other servers in zero-knowledge
that it knows the private keys that match the public keys.
All public keys are made available to all servers and users.

\subsection{Sending messages with AHS} \label{sec:active_clients}
Once the servers' keys are distributed, user Alice can submit
a message to a chain. To do so, Alice now employs
a double-enveloping technique to encrypt her message~\cite{golle2002}:
she first onion-encrypts her message for all servers using the inner keys,
and then onion-encrypts the result with the mixing keys.
Let \emph{inner ciphertext} be the result of the first onion-encryption,
and \emph{outer ciphertext} be the final ciphertext.
The inner ciphertexts are encrypted using $\prod_i \innerkey_i$
as the public key, which allows users to onion-encrypt in ``one-shot'':
i.e., $\inner = (g^y, \aenc(\kex(\prod_i \innerkey_i, y), \round, \msg))$
in round $\round$ with message $\msg$ and a random $y$
(Without $y$, $\kex(\prod_i \innerkey_i, y)$
can only be computed if all $\{\privateinnerkey_i\}$ are known).
To generate the outer ciphertext, Alice performs the following.
\begin{enumerate}
  \item Generate her outer Diffie-Hellman key: a random $x \in \Z_p$ and $(g^x, x)$.
  \item \label{step:client_nizk}
    Generate a NIZK that proves she knows $x$ that matches $g^x$
    (using knowledge of discrete log proof~\cite{camenisch-log}).
  \item \label{step:client_submit}
    Let $\outerc_{\chain+1} = \inner$, and
    let $\{\mixingkey_i\}$ for $i \in [\chain]$ be the mixing keys of
    the servers in the chain.
    For $i = \chain$ to 1, compute
    $\outerc_i = \aenc(\kex(\mixingkey_i, x), \round, \outerc_{i+1})$.
\end{enumerate}
$\outerc = (g^x, \outerc_1)$ is the final outer ciphertext.
This is nearly identical to Algorithm~\ref{alg:client},
except that the user does not generate a fresh pair of Diffie-Hellman
keys for each layer of encryption.
To submit the message, Alice sends $\outerc$ and the NIZK to
all servers in the chain.

\subsection{Mixing with AHS} \label{sec:active_mixing}
Before mixing begins in round $\round$, the servers have
${\{c^j = (X^j_1 = g^{x_j}, \outerc_1 = \aenc(\kex(\mixingkey_1, x_j), \round, \outerc_2^j))\}}$.
The servers first agree on the inputs for this round.
This can be done, for example, by sorting the users' ciphertexts,
hashing them using a cryptographic hash function, and then comparing the hashes.
Then, starting with server 1, server $i=1,\ldots,\chain$ perform the following:
\begin{enumerate}
  \item \textbf{Decrypt and shuffle:}
    Similar to Algorithm~\ref{alg:mix}, decrypt each message.
    Each message is of the form
    $(X^j_i, \outerc^j_i = \aenc(\kex(\mixingkey_i, x_j), \round, \outerc_{i+1}^j)$.
    Thus,
    ${(b^j, \outerc_{i+1}^j) = \adec(\kex(X^j_i, \privatemixingkey_i), \round, \outerc_i^j)}$.
    If any decryption fails (i.e., $b^j = 0$ for some $j$),
    then mixing halts and the server can start a blame protocol described
    in \S\ref{sec:blame}.
    Randomly shuffle $\{\outerc_{i+1}^j\}$.
  \item \textbf{Blind and shuffle:}
    Blind the users' Diffie-Hellman keys $\{X^j_i\}$:
    $X^j_{i+1} = (X^j_i)^{\privateblindkey_i}$ for each $j$.
    Then, shuffle the keys using the same permutation
    as the one used to shuffle the ciphertexts.
  \item \textbf{Generate zero-knowledge proof:} \label{step:server_nizk}
    Generate a proof that
    $(\prod_j X^j_i)^{\privateblindkey_i} = \prod_j X^j_{i+1}$
    by generating a NIZK that shows
    ${\log_{\prod_j (X^j_i)}(\prod_j X^j_{i+1}) = \log_{\blindkey_{i-1}} \blindkey_i} (= \privateblindkey_i)$.
    Send the NIZK with the shuffled $\{X^j_{i+1}\}$
    to all other servers in the chain.
    All other servers verify this proof is valid using the $\{X^j_i\}$
    they received previously,
    $\{X^j_{i+1}\}$, $\blindkey_{i-1}$, and $\blindkey_i$.
  \item \textbf{Forward messages:} \label{step:active_forward}
     If $i < \chain$, then send the shuffled $\{(X^j_{i+1}, \outerc_{i+1}^j)\}$
    to server $i+1$.
\end{enumerate}
When the last server finishes shuffling and
no server reports any errors during mixing,
our protocol guarantees that the honest server mixed all
the honest users' messages successfully,
meaning the users' privacy was protected. At this point,
the servers reveal their private per-round inner keys $\{\privateinnerkey_i\}$.
With this, the last server can decrypt the inner ciphertexts
to recover the users' messages.

To demonstrate the correctness of AHS
(i.e., every message is successfully delivered if
every participant followed the protocol),
consider the key user $j$ used to encrypt the message intended for server $i$
and the Diffie-Hellman key server $i$ receives.
User $j$ encrypts the message using the key
$\kex(\mixingkey_i, x_j) = g^{x_j \cdot \privatemixingkey_i \prod_{a<i} \privateblindkey_a}$.
The Diffie-Hellman key server $i$ receives is
$X^j_i = g^{x_j \cdot \prod_{a<i} \privateblindkey_a}$.
The key exchange then results in
$\kex(X^j_i, \privatemixingkey_i) = g^{\privatemixingkey_i \cdot x_j \prod_{a<i} \privateblindkey_a}$,
which is the same as the one the user used.

\paragraph{Analysis.}
We now provide a brief security analysis of our scheme.
This scheme provides protection against honest-but-curious adversaries
as the inputs and outputs of an honest server look random:
If decisional Diffie-Hellman is hard,
then $g^{x \cdot \privateblindkey}$ is indistinguishable from a random value
given $g^{x}$ and $g^\privateblindkey$ for random $x$ and $\privateblindkey$.
Thus, by observing $\{g^{x_j}\}$ (input) and
$\{g^{x_{\pi(j)} \cdot \privateblindkey}\}$ (output) of an honest server
where $\pi$ is the random permutation used to shuffle the messages,
the adversary does not learn anything about the relationships between
the inputs and outputs.

We now provide a high level analysis that
the honest server will always detect tampering
by an upstream adversary.
The detailed proof is shown in Appendix~\ref{app:active_security}.
Let server $h$ be the honest server.
First, since we only need to consider upstream adversaries,
we will consider all upstream malicious servers as one collective server with
private blinding key $\privateblindkey_A = \sum_{i < h} \privateblindkey_i$.
For the adversary to successfully tamper,
it must generate $\{X^j_h\}$ such that
${(\prod X^j_1)^{\privateblindkey_A} = \prod X^j_h}$
(otherwise it would fail the NIZK verification in
step~\ref{step:server_nizk} of \S\ref{sec:active_mixing}).
Let $X_T \neq \emptyset$ be the set of honest users whose messages were tampered.
The adversary cannot compute
$((X^j_1)^{\privateblindkey_A})^{\privatemixingkey_h}$ for $j \in X_T$,
the keys the users used to authenticate the messages,
since Diffie-Hellman is hard.
Thus, the adversary must change the users' Diffie-Hellman keys
in order to generate authenticated ciphertexts that differ from the original ciphertexts;
i.e., $(X^j_1)^{\privateblindkey_A} \neq X^j_h$ for $j \in X_T$.
Moreover, because all decryption operations must be successful
to avoid detection,
the adversary must know the keys used for authenticated decryption,
which are $(X^j_h)^{\privatemixingkey_h}$ for $j \in X_T$.

In the beginning of a round,
the adversary controlled users have to prove the knowledge of discrete log
of their Diffie-Hellman keys after seeing
$\{X^j_1 = g^{x_j}\}$ from honest users.
Then, the adversarial users' keys are generated independently
of the honest users'.
As a result, the goal of the adversary is essentially to find
$\{X^j_h\}_{j \in X_T}$ such that
${(\prod_{j \in X_T} X^j_1)^{\privateblindkey_A} = \prod_{j \in X_T} X^j_h}$,
with $(X^j_1)^{\privateblindkey_A} \neq X^j_h$.
However, if the adversary could do so,
then it could also compute
${((\prod_{j \in X_T} X^j_1)^{\privateblindkey_A})^{\privatemixingkey_h}
  = \prod_{j \in X_T} (X^j_h)^{\privateblindkey_h}}$,
since it knows $(X^j_h)^{\privatemixingkey_h}$.
This means that the adversary computed
${((\prod_{j \in X_T} X^j_1)^{\privateblindkey_A})^{\privatemixingkey_h}
= g^{\privatemixingkey_h \cdot \privateblindkey_A \cdot \sum_{j \in X_T} x_j}}$
only given $\{g^{x_j}\}_{j \in X_T}$, $\privateblindkey_A$,
and $(g^{\privateblindkey_A})^{\privatemixingkey_h}$,
where $\{x_j\}_{j \in X_T}$ and $\privatemixingkey_h$ are random values
independent of $\privateblindkey_A$.
This breaks the Diffie-Hellman assumption,
and thus the adversary must not be able to tamper with messages undetected.

\subsection{Blame protocol} \label{sec:blame}
There are two ways an adversary could be detected:
a NIZK fails to verify or an authenticated decryption fails.
If a malicious user cannot generate a correct NIZK in
step~\ref{step:client_nizk} in \S\ref{sec:active_clients}
or if a malicious server misbehaves and cannot generate a correct NIZK in
step~\ref{step:server_nizk} in \S\ref{sec:active_mixing},
then the misbehavior is detected
and the adversary is immediately identified.
In the case where a server finds some misauthenticated ciphertexts,
the server can start a blame
protocol that allows the server to identify who misbehaved.
At a high level, the protocol guarantees that users are identified
if and only if they purposefully sent misauthenticated ciphertexts.
In addition, the protocol ensures that the privacy of honest users
always remain protected,
even if a malicious server tries to falsely accuse honest users.

Once server $h$ identifies an misauthenticated ciphertext,
it starts the blame protocol by revealing
the problem ciphertext $(X^j_h, \outerc^j_h)$.
Then, the servers execute the following:
\begin{enumerate}
  \item \label{step:blame_keys}
    For $i = h-1, \ldots, 1$, the servers reveal in order $X^j_i$
    that matches $X^j_{i+1}$ (i.e., $(X^j_i)^{\privateblindkey_i} = X^j_{i+1})$.
    Each server proves to all other servers it calculated
    $X^j_i$ correctly by showing that
    ${\log_{X^j_i}(X^j_{i+1}) = \log_{\blindkey_{i-1}}(\blindkey_i) (= \privateblindkey_i})$.
  \item \label{step:blame_ciphertext}
    For $i = h-1, \ldots, 1$, the servers reveal in order $\outerc^j_i$
    that matches $\outerc^j_{i+1}$ (i.e., $\outerc^j_i = \aenc(\kex(X^j_i, \privatemixingkey_i), \round, \outerc^j_{i+1})$.
    Each server proves it correctly decrypted the ciphertext by
    revealing the key used for decryption  $(X^j_i)^{\privatemixingkey_i}$,
    and showing that
    ${\log_{X^j_i}((X^j_i)^{\privatemixingkey_i}) = \log_{\blindkey_{i-1}}(\mixingkey_i)}$.
    The other servers can verify the correctness of the decryption operation
    by checking the NIZK and decrypting the ciphertext themselves.
  \item All servers check that $\outerc^j_1$ revealed by the first server
    matches the user submitted ciphertext
    (from \S\ref{sec:active_clients}).
  \item Similar to step~\ref{step:blame_ciphertext},
    server $h$ (the accusing server) reveals its Diffie-Hellman exchanged key
    $(X^j_h)^{\privatemixingkey_h}$,
    and shows that
    ${\log_{X^j_h}((X^j_h)^{\privatemixingkey_h}) = \log_{\blindkey_{h-1}}(\mixingkey_h)}$.
    All servers verify that $\adec(k_h, \outerc^j_h)$ fails.
\end{enumerate}
If there are multiple problem ciphertexts,
the blame protocol can be carried out in parallel for each ciphertext.
Steps~\ref{step:blame_keys} and~\ref{step:blame_ciphertext}
can be done simultaneously as well.
If the servers successfully carry out the blame protocol,
then the servers have identified actively malicious users.
At this point, those ciphertexts are removed from the set,
and the upstream servers are required to repeat the AHS protocol;
since the accusing servers have already removed all bad ciphertexts,
the servers just have to
repeat step~\ref{step:server_nizk} of \S\ref{sec:active_mixing}
to show the keys were correctly computed.
If any of the above steps fail, then the servers delete their private inner keys.

\paragraph{Analysis.}
The accusing server and the upstream servers are
required to reveal the exchanged key used to decrypt the ciphertexts,
and the correctness of the key exchange is proven through the two NIZKs in
step~\ref{step:blame_keys} and step~\ref{step:blame_ciphertext}.
Then, all servers (in particular, the honest server) can use the revealed keys
to ensure that the submitted original ciphertext decrypts to the problem ciphertext.
Thus, if a user submits misauthenticated ciphertext,
then the servers get a verifiable chain of decryption
starting with the outer ciphertext to the problem ciphertext,
allowing the servers to identify the malicious user.
(Intuitively, the outer ciphertext behaves
as a commitment for the all layers of encryption.)
Moreover, if an honest user submits a correctly authenticated ciphertext,
she will never be accused successfully,
since an honest user's ciphertext will authenticate at all layers.
Thus, if a user is actively malicious, then she will be identified and removed.

Finally, even after a false accusation, the users' privacy is protected.
After a malicious server accuses an honest user,
the malicious server learns that either
an outer ciphertext (if the server is upstream of server $h$)
or an inner ciphertext (if the server is downstream of server $h$)
originated from the user.
In either case, the message remains encrypted for the honest server,
by the mixing key in the former case or by the inner key in the latter case.
The blame protocol will fail when the malicious server fails to prove
that the honest user's ciphertext is misauthenticated,
and the ciphertext will never be decrypted.
As such, the adversary never learns the final destination of the user's message,
and \sys protects the honest users' privacy.

\section{Implementation} \label{sec:impl}
The \sys prototype is written in approximately 3,500 lines of Go.
We used NaCl~\cite{nacl} for authenticated encryption,
which internally uses ChaCha20 and Poly1305.
The servers communicated using streaming gRPC over TLS.
Our prototype assumes that the servers' and users' public keys,
and the public randomness for creating the chains are provided
by a higher level functionality.
Finally, we set the number of chains $\chains$ equal
to the number of servers $\servers$ (\S\ref{sec:anytrust}).

\section{Evaluation} \label{sec:eval}
In this section, we investigate the cost of users
and the performance of \sys
for different network configurations.
For majority of our experiments, we assumed $f=0.2$
(i.e., 80\% of the servers are honest)
unless indicated otherwise.
We used 256~byte messages,
similar to evaluations of prior systems~\cite{vuvuzela,stadium,pung};
this is about the size of a standard SMS message or a Tweet.
We used c4.8xlarge instances on Amazon EC2 for our experiments,
which has 36 Intel Xeon E5-2666 CPUs
with 60~GB of memory and 10~Gbps links.

We compare the results against three prior systems:
Stadium~\cite{stadium}\footnote{We
  did not compare against Karaoke~\cite{karaoke}
  which is significantly faster than Stadium~\cite{stadium}
  since the implementation is not yet available.}
Atom~\cite{atom}, and Pung~\cite{pung,sealpir}.
For Stadium, we report the performance for $e^\epsilon = 10$
(meaning the probability of Alice talking with Bob is within $10\times$
the probability of Alice talking with any other user)
and allow up to 10,000 rounds of communication
with strong security guarantees ($\delta < 10^{-4}$),
which are the parameters the used for evaluation in their paper.
We show results for Pung with XPIR (used in the original paper~\cite{pung})
and with SealPIR (used in the follow-up work~\cite{sealpir})
when measuring user overheads,
and only with XPIR when measuring end-to-end latency.
We do this because SealPIR
significantly improves the client performance
and enables more efficient batch messaging,
but introduces extra server overheads in the case of one-to-one messaging.

As mentioned in \S\ref{sec:related},
the three systems scale horizontally,
but offer different security properties from \sys.
To summarize, Stadium provides
differential privacy guarantees
against the same adversary assumed by \sys.
Stadium users can exchange up to 10,000 sensitive messages
with strong privacy,
while \sys users can do so for unlimited messages.
Atom provides cryptographic sender anonymity~\cite{pfitzmann2010terminology}
under the same threat model.
Finally, Pung provides messaging with cryptographic privacy against
a stronger adversary who can compromise all servers
rather than a fraction of servers.
To the best of our knowledge, we are not aware
of any scalable private messaging systems
that offer the same security guarantee under a similar threat model to \sys.

\subsection{User costs} \label{sec:user_costs}
\begin{figure}[t]
  \centering
  \includegraphics[width=\linewidth]{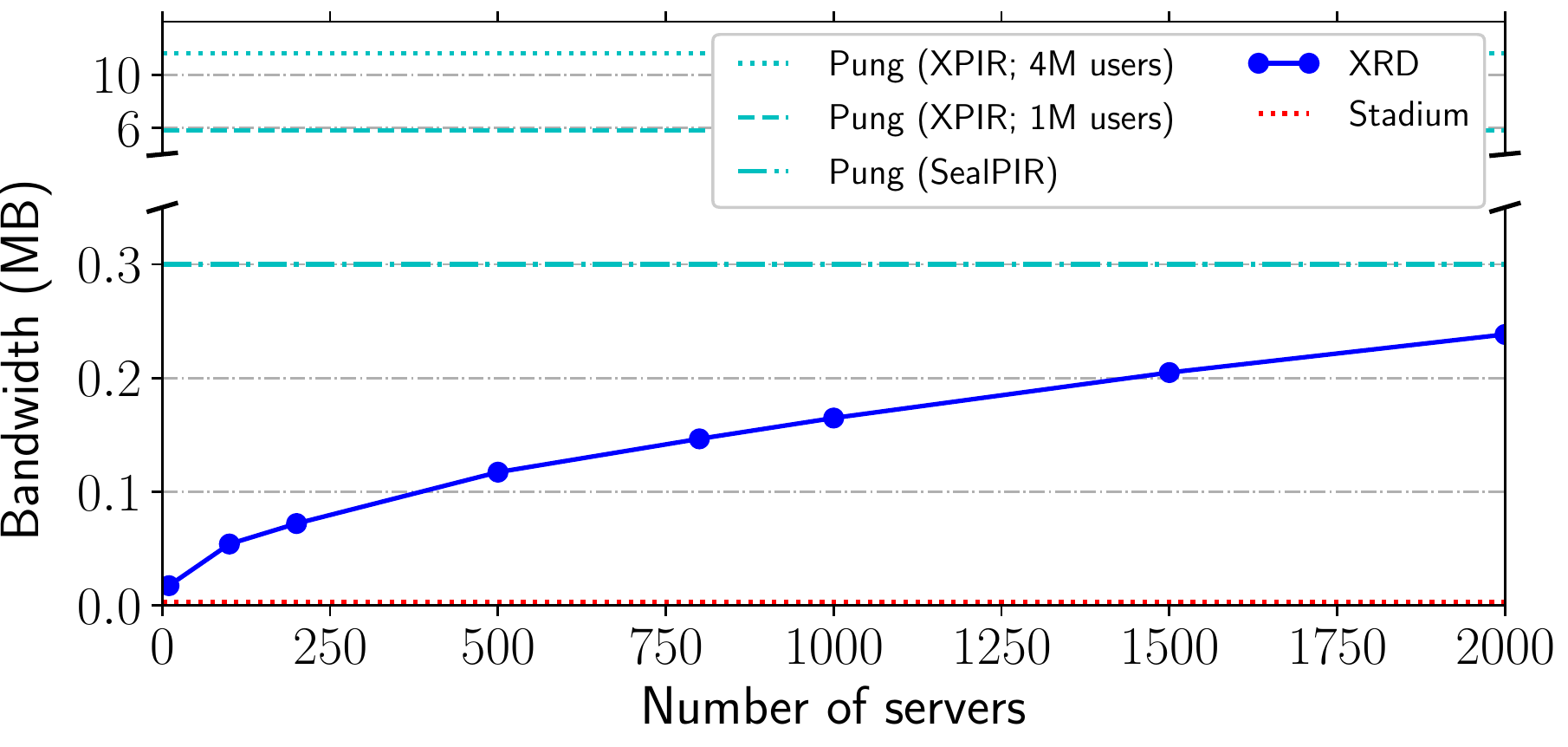}
  \caption{Required user bandwidth per round as a function
    of number of servers in the network.}
  \label{fig:user_bandwidth}
\end{figure}

\begin{figure}[t]
  \centering
  \includegraphics[width=\linewidth]{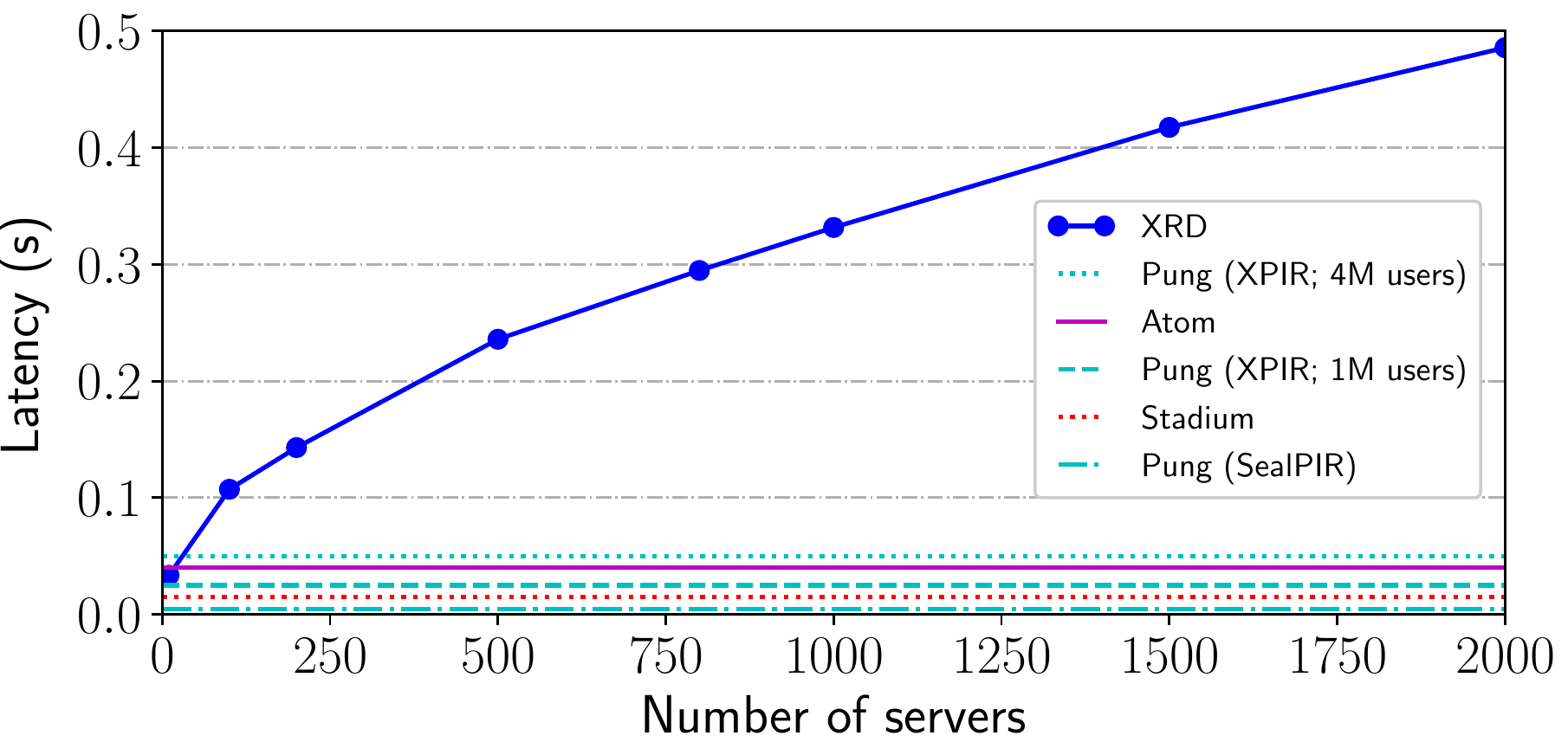}
  \caption{Required user computation as a function
    of number of servers with a single core.
    The computation could easily be parallelized
    with more cores for \sys.
  }
  \label{fig:user_latency}
\end{figure}

We first characterize computation and bandwidth overheads of \sys users
using a single core of a c4.8xlarge instance.
In order to ensure that every pair of users intersects,
each user sends $\sqrt{2\servers}$ messages (\S\ref{sec:chain_selection}).
This means that the overheads for users increase as we add more servers
to the network, as shown in Figure~\ref{fig:user_bandwidth}
and~\ref{fig:user_latency}.
This is a primary weakness of \sys,
since our horizontal scalability comes at a higher cost for users.
Still, the cost remains reasonable even for large numbers of servers.
With 100 \sys servers, each user must submit about 54~KB of data.
The bandwidth usage increases to about 238~KB of data with 2,000 servers.
For 1 minute rounds,
this translates to about 40~Kbps of bandwidth requirement.
A similar trend exists for computation overhead as well,
though it also remains relatively small:
it takes less than 0.5 seconds with fewer than 2,000 servers in the network.
This could be easily parallelized with more cores,
since the messages for different chains are independently generated.
The cover messages make up half of the client overhead
(\S\ref{sec:user_churn}).

User costs in prior works
do not increase with the number of servers.
Still, Pung with XPIR incurs heavy user bandwidth overheads
due to the cost of PIR.
With 1 million users, Pung users transmit about 5.8~MB,
which is about $25\times$ worse than \sys
when there are fewer than 2,000 servers.
Moreover, per user cost of XPIR is proportional
to the total number of users ($O(\sqrt{\users})$ for $\users$ users):
the bandwidth cost increases to 11~MB of bandwidth
for 4 million users.
The SealPIR variant, however, is comparable to that of \sys,
as the users can compress the communication
using cryptographic techniques.
Finally, Stadium and Atom incur minimal user bandwidth cost,
with less than a kilobyte of bandwidth overhead.
Thus, for users with heavily limited resources
in a large network,
prior works can be more desirable than \sys.
Lowering the user bandwidth cost in \sys so that it is comparable
to prior systems is an interesting future direction.

\subsection{End-to-end latency} \label{sec:end2end}
\paragraph{Experiment setup.}
To evaluate the end-to-end performance,
we created a testbed consisting of up to 200 c4.8xlarge instances.
We ran the instances within the same data center
to avoid bandwidth costs,
but added 40-100ms of round trip latency
between servers using the Linux \texttt{tc} command
to simulate a more realistic distributed network.

We used many c4.8xlarge instances to simulate millions of users,
and also used ten more c4.8xlarge instances to simulate
the mailboxes. We generate all users' messages before
the round starts, and measure the critical path of our system
by measuring the time between
the last user submitting her message
and the last user downloading her message.

We estimate the latency of Pung
with $\users$ users and $\servers$ servers
by evaluating it on a single c4.8xlarge instance
with $\users/\servers$ users.
This is the best possible latency
Pung can achieve because
(1) Pung is embarrassingly parallel, so evenly dividing
users across all the servers should be ideal~\cite[\S7.3]{pung},
and (2) we are ignoring the extra work needed for coordination
between the servers (e.g., for message replication).
For Stadium, we report the latency when the length
of each mix chain is nine servers.

We focus on the following questions in this section,
and compare against prior work:
\begin{itemize}
  \item What is the end-to-end latency of \sys,
    and how does it change with the number of users?
  \item How does \sys scale with more servers?
  \item What is the effect of $f$, the fraction of malicious servers,
    on latency?
  \item What is the impact of the blame protocol on performance?
\end{itemize}

\begin{figure}[t]
  \centering
  \includegraphics[width=\linewidth]{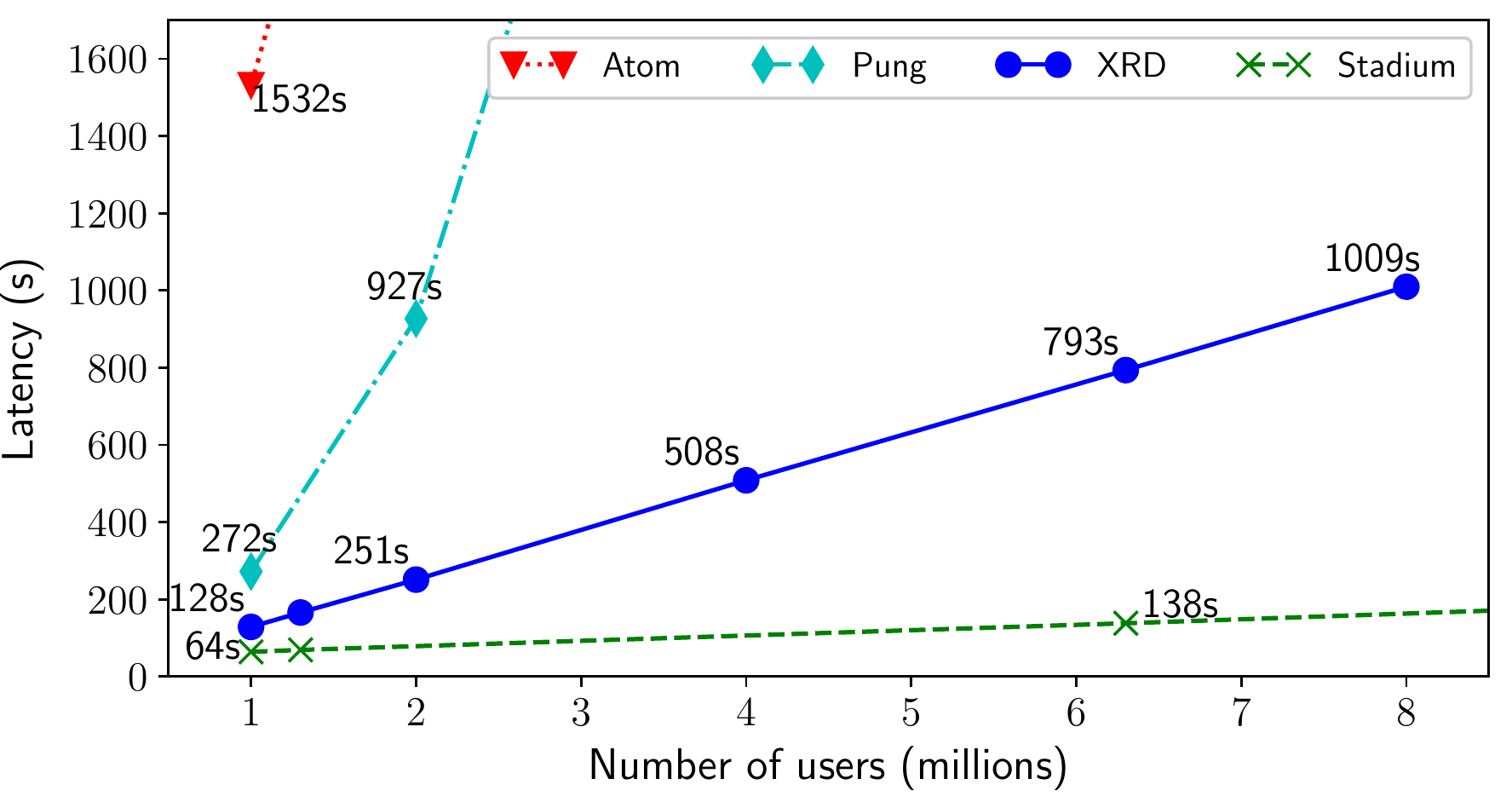}
  \caption{End-to-end latency of \sys and prior systems
    with varying numbers of users with 100 servers.
  }
  \label{fig:users_vs_latency}
\end{figure}

\paragraph{Number of users.}
Figure~\ref{fig:users_vs_latency} shows the end-to-end latency
of \sys and prior works with 100 servers.
\sys was able to handle 1 million users in 128 seconds,
and the latency scales linearly with the number of users.
This is $12\times$ and $2.1\times$ faster than Atom and Pung,
and $2\times$ worse than Stadium
for the same deployment scenario.
Though processing a single message in \sys is significantly
faster than doing so in Stadium (since Stadium relies on
verifiable shuffle, while \sys uses AHS),
the overall system is still slower.
This is because each \sys user submits many messages.
For example, each user submits 15 messages with 100 servers,
which is almost equivalent to adding 15 users who each submit one message.
Unfortunately, the performance gap would grow with more servers
due to each user submitting more messages
(the rate at which the gap grows would be proportional
to $\sqrt{2\servers}$).
While \sys cannot provide the same performance as Stadium
with large numbers of users and servers,
\sys can provide stronger cryptographic privacy.

When compared to Pung,
the speed-up increases further with the number of users
since the latency of Pung grows superlinearly.
This is because
the server computation per user increases
with the number of users.
With 2 and 4 million users, for example, \sys is $3.7\times$ and $7.1\times$ faster.
For Atom, the latency increases linearly,
but the performance gap still increases with more users.
This is due to its heavy reliance on expensive public key cryptography
and long routes for the messages (over 300 servers).

\begin{figure}[t]
  \centering
  \includegraphics[width=\linewidth]{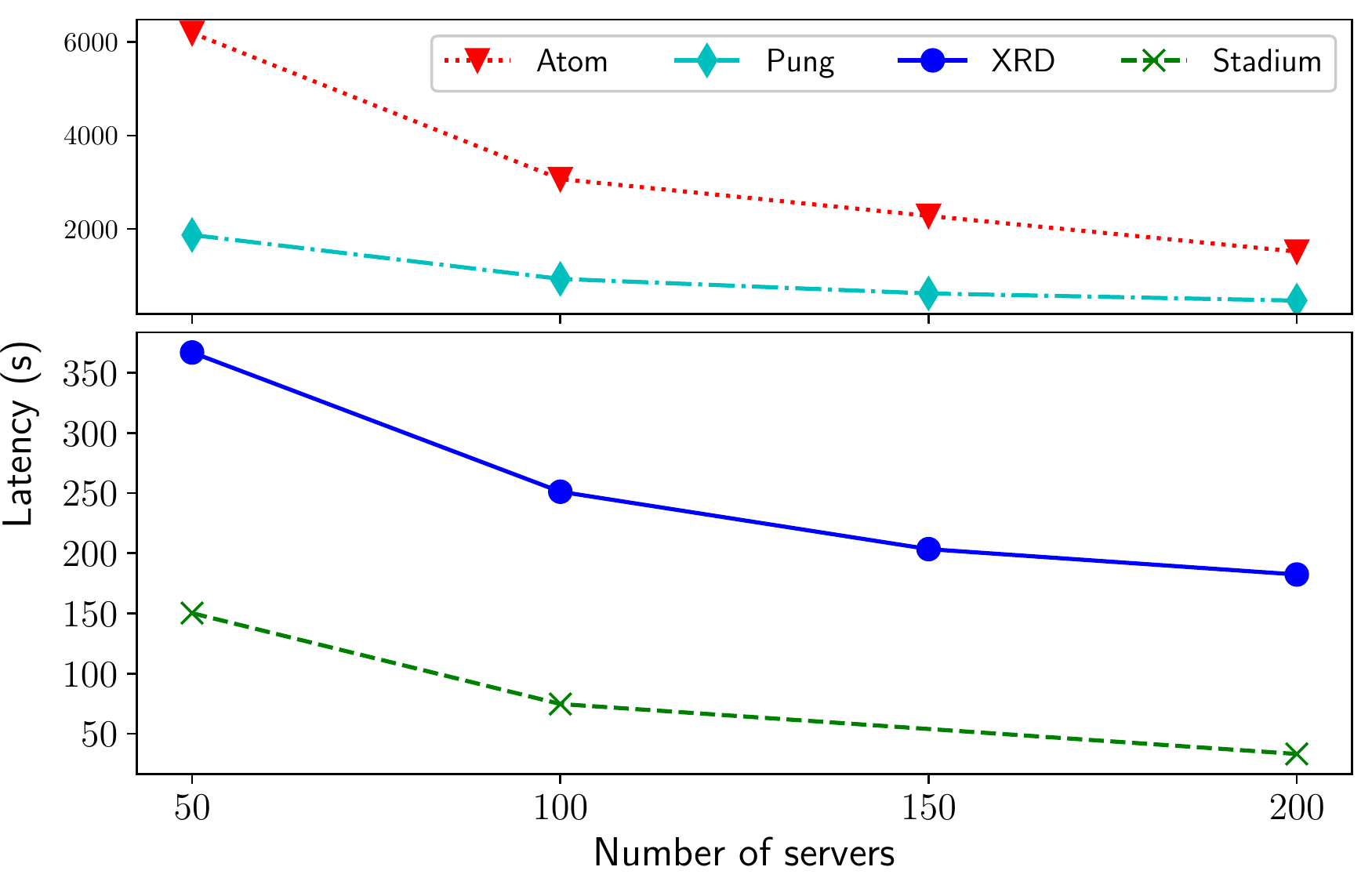}
  \caption{End-to-end latency of \sys for varying numbers of servers
      with 2 million users.
      We show Pung and Atom on a different time scale.}
  \label{fig:servers_vs_latency}
\end{figure}

\paragraph{Scalability.}
Figure~\ref{fig:servers_vs_latency} shows how the latency
decreases with the number of servers with 2 million users.
We experimented with up to 200 servers,
and observed the expected scaling pattern:
the latency of \sys reduces as $\sqrt{2/\servers}$
with $\servers$ servers (\S\ref{sec:scalability}).
In contrast, prior works scale as $1/\servers$,
and thus will outperform \sys with enough servers.
Still, because \sys employs more efficient cryptography,
\sys outperforms Atom and Pung with less than 200 servers.

To estimate the performance of larger deployments,
we extrapolated our results to more servers.
We estimate that \sys can support 2 million users with 1,000 servers
in about 84s, while Stadium can do so in about 8s.
(At this point, the latency between servers would be the dominating factor for Stadium.)
This gap increases with more users, as described previously.
For Atom and Pung, we estimate that the latency would
be comparable to \sys with about 3,000 servers
and 1,000 servers in the network, respectively, for 2 million users.
Pung would need more servers with more users
to catch up to \sys due to the superlinear increase in latency
with the number of users.

\begin{figure}[t]
  \centering
  \includegraphics[width=\linewidth]{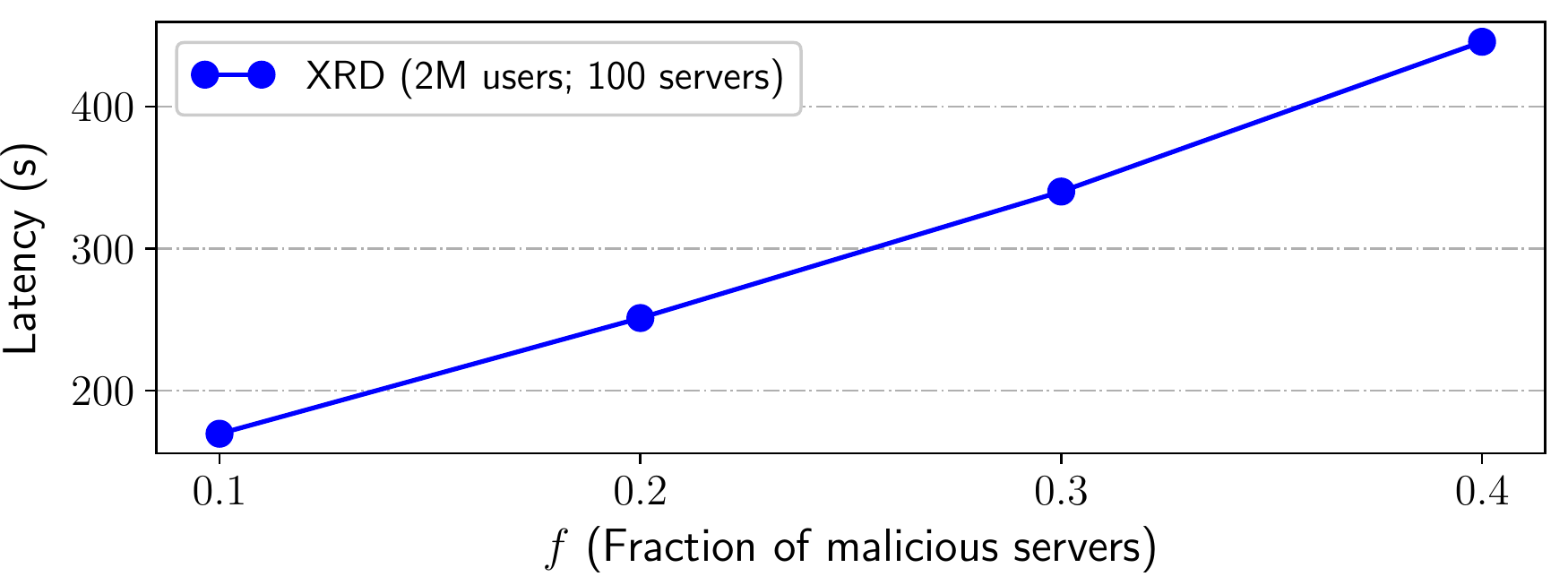}
  \caption{Latency of \sys for different values of $f$.}
  \label{fig:f_vs_latency}
\end{figure}

\paragraph{Impact of $f$.}
During setup, the system administrator should make a conservative
estimate of the fraction of malicious servers $f$ to form the groups.
Larger $f$ affects latency because it increases in
the length of the chains $\chain$ (\S\ref{sec:anytrust}).
Concretely, with $\chains = 100$, $\chain$ must satisfy
$100 \cdot f^\chain < 2^{-64}$. Thus, $\chain > \frac{\log(2^{-64}/100)}{\log(f)}$,
which means that
the length of a chain (and the latency) grows as a function of $\frac{-1}{\log(f)}$.
Figure~\ref{fig:f_vs_latency} demonstrates
this effect.
This function grows rapidly when $f >> 0.5$,
and thus the latency would be significantly worse
when considering larger values of $f$.
Atom would experience the same effect
since its mix chains are created using the same strategy as \sys.
Stadium would face more significant increase in latency with $f$
as its mix chains also similarly get longer with $f$,
and the length of the chains has a superlinear
effect on the latency due to
expensive zero-knowledge proof verification~\cite[\S10.3]{stadium}.
The latency of Pung does not increase with $f$ since it already assumes $f=1$.

\begin{figure}[t]
  \centering
  \includegraphics[width=\linewidth]{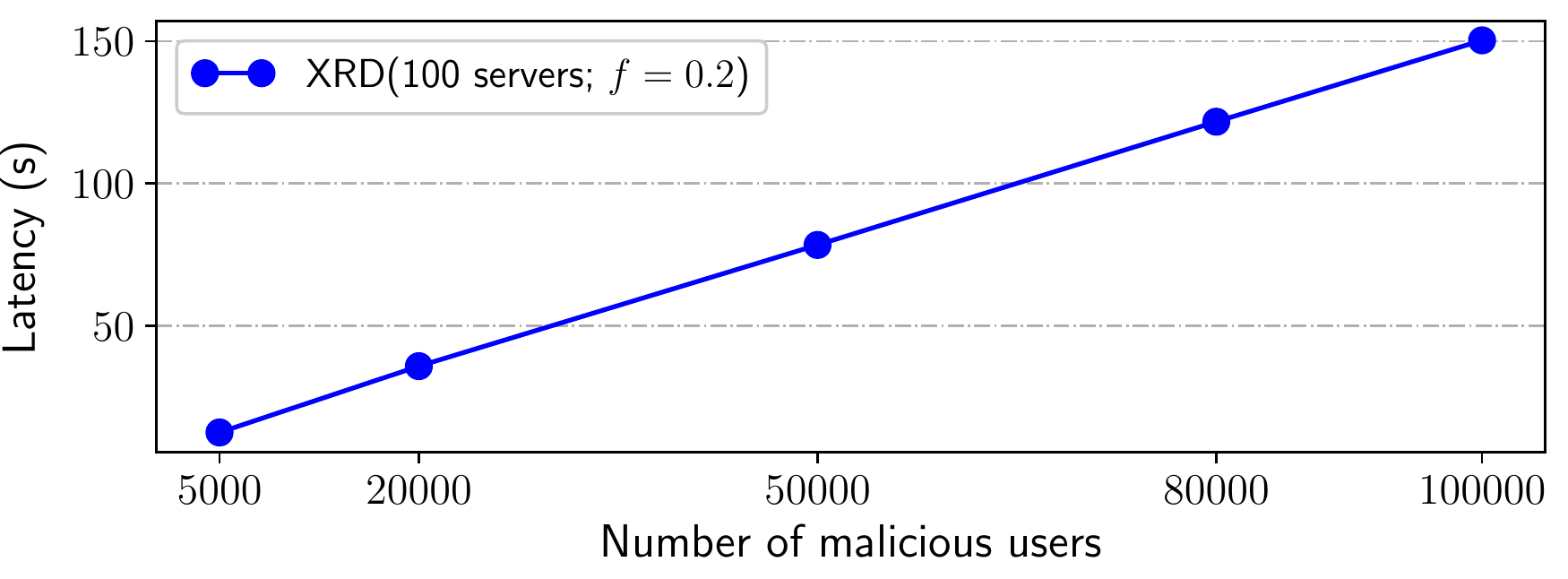}
  \caption{Latency of blame protocol}
  \label{fig:blame_latency}
\end{figure}

\paragraph{Impact of blame protocol.}
The performance shown in Figure~\ref{fig:users_vs_latency}
and Figure~\ref{fig:servers_vs_latency}
assume that no misbehavior was detected.
While malicious users by themselves cannot deny service,
they could send misauthenticated ciphertexts in an attempt to
trigger the blame protocol, and slow down the protocol.
Since malicious users are removed from the network when detected,
they cause the most slowdown when the misauthenticated ciphertexts
are at the last server.
The performance of blame protocol also depends on how many
users were caught this round.
We therefore show the worst case latency increase due to blame protocol
as a function of number of malicious users in a chain
in Figure~\ref{fig:blame_latency} with $f=0.2$.
The blame protocol requires two discrete log equivalence proof
and decryption per user for each layer of encryption.
Concretely, if 5,000 users misbehave in a chain,
the blame protocol takes about 13 seconds to finish.
This cost increases linearly with the number of users.
For example, if 100,000 users misbehave in a chain
blame protocol takes about 150 seconds.
(The later case, for example, corresponds to
when a third of all users are malicious
with 100 servers and 2 million users in the network.)
While this is a significant increase in latency,
malicious users can be removed from the network once malicious users are detected.
Thus, to cause serious slowdowns across many rounds, the adversary needs
large amounts of resources in order to constantly create new malicious users.
Moreover, \sys remains at least $6\times$ and $2\times$
faster than Atom and Pung even with
100,000 malicious users in a chain.
The faster version of Atom used for comparison in this paper
(the trap message variant) does not protect against malicious user
denial-of-service.
Protecting against adversarial users would require at least
$4\times$ slowdown for Atom~\cite[\S6]{atom}.

\subsection{Availability} \label{sec:availability}
\begin{figure}[t]
  \centering
  \includegraphics[width=\linewidth]{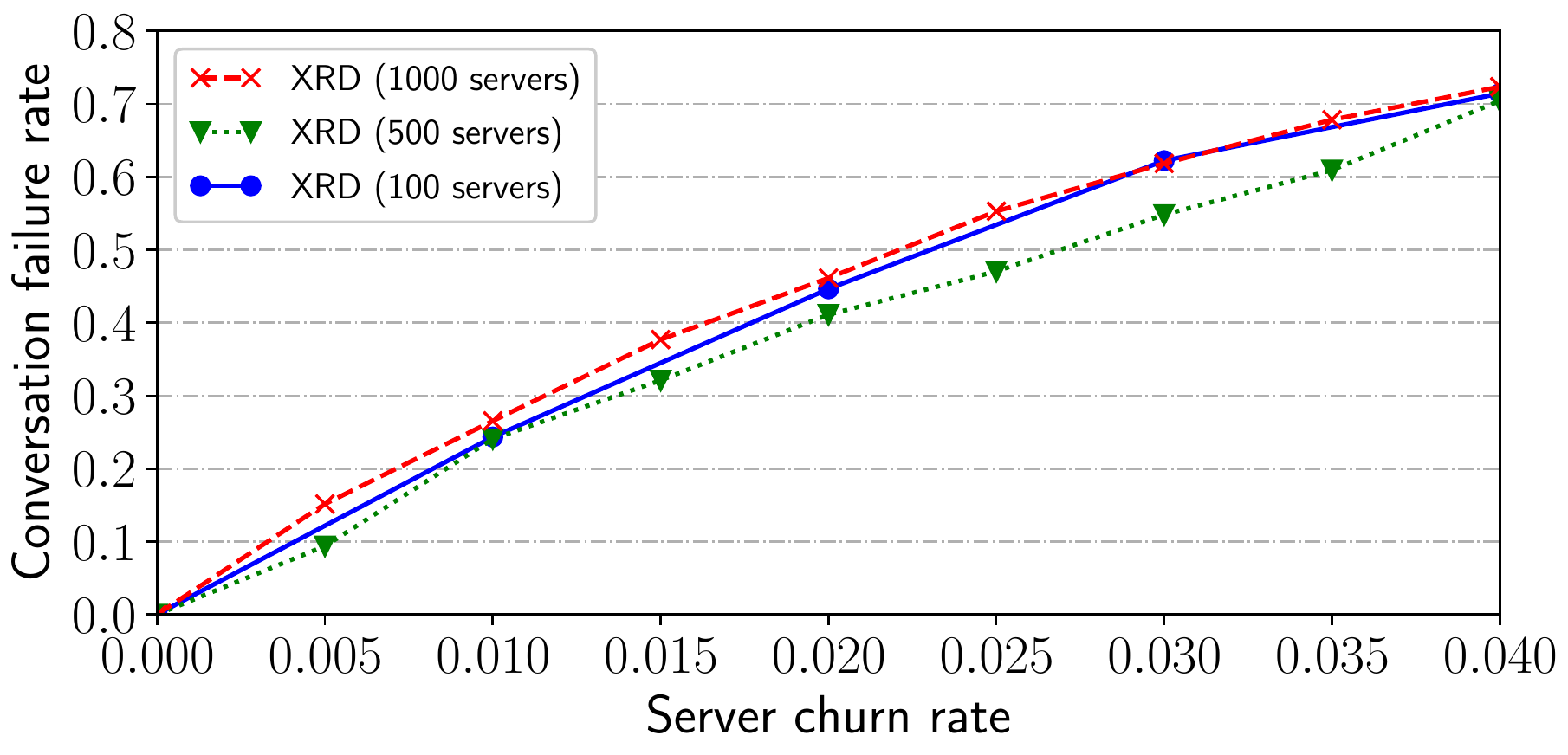}
  \caption{Fraction of conversations that fail in
    a given round due to server failures
    for different server churn rates.
  }
  \label{fig:availability}
\end{figure}

To estimate the effect of server churn on a \sys network,
we simulated deployment scenarios with 2 million users
and different numbers of servers.
We assumed that all users were in a conversation,
and show the fraction of the users whose conversation
messages did not reach their partner in Figure~\ref{fig:availability}.
For example, if 1\% of the servers fail in a given round
(comparable to server churn rate in Tor~\cite{tor_metric}),
then we expect about 27\% of the conversations to experience failure,
and the end-points would have to resend their conversation messages.
Unfortunately, the failure rate quickly increases with the server
churn rates, reaching 70\% with 4\% server failures,
as more chains contain at least one failing server.
Thus, it would be easy for the adversary who controls
a non-trivial fraction of the servers to launch a denial-of-service attack.
Addressing this concern remains important future work.

When compared to Pung, the availability guarantees can be significantly worse,
assuming Pung replicates all users' messages across all servers.
In this case,
the server churn rate would be equal to the user failure rate
with users evenly distributed across all Pung servers,
and the users connected to failing servers could be rerouted to
other servers to continue communication.
Atom can tolerate any fraction $\churn$ of the servers failing
using threshold cryptography~\cite{threshold},
but the latency grows with $\churn$.
For example, to tolerate $\churn = 1\%$ servers failing,
we estimate that Atom would be about 10\% slower~\cite[Appendix~B]{atom}.
Finally, Stadium uses two layers of parallel mix-nets
(i.e., each layer is similar to \sys),
and fully connects the chains across the two layers.
As a result, even one server failure would cause
the whole system to come to a halt.
(Stadium does not provide a fault recovery mechanism,
and the security implications of continuing
the protocol without the failing chains are not analyzed~\cite{stadium}.)

\section{Discussion and future work} \label{sec:discuss}

\paragraph{Multi-user conversations.}
In this paper, we focused on providing private one-to-one conversations.
A natural extension would be to provide group conversations.
\sys can already provide group conversations in scenarios
where the users in a group chat intersect at different chains.
For example, consider three users Alice, Bob, and Charlie who wish
to have a private group conversation.
If (Alice, Bob), (Alice, Charlie), and (Bob, Charlie) all intersect
at different chains, then each user could carry out one-to-one conversation
on two different chains to have a group conversation.
The same mechanism can be used to have multiple one-to-one conversations as well.
This could help amortize the cost per conversation
since more than one of $\sqrt{2n}$ messages would carry conversation messages.
However, \sys currently cannot support multiple conversations for one user
if she intersects with different partners at the same chain
(e.g., Alice intersects with Bob and Charlie on the same chain).
We wish to generalize \sys to multi-user conversations in the future.

\paragraph{Values of $\chainsperuser$ and load distribution.}
As discussed in \S\ref{sec:overview},
the number of chains each user selects
is $\chainsperuser \geq \sqrt{\servers}$,
and our chain selection algorithm in \S\ref{sec:chain_selection}
is a $\sqrt{2}$-approximation algorithm.
It maybe possible, however, to find a better algorithm
that yields lower $\chainsperuser$, closer to $\sqrt{\servers}$.
This would result in proportional speed-up of \sys,
up to $\sqrt{2}\times$ faster.
Furthermore, since \sys only requires that all users
intersect, we need not use the same $\chainsperuser$ for all users,
or evenly distribute the load across all chains.
If different users and chains have different capacities,
it may be beneficial for performance to vary $\chainsperuser$ per user
with uneven distribution of messages.

\section{Conclusion} \label{sec:conc}
\sys provides a unique design point in the space of
metadata private communication systems by achieving
cryptographic privacy and horizontal scalability
using efficient cryptographic primitives.
\sys organizes the servers into multiple small chains
that process messages in parallel,
and can scale easily with the number of servers by adding more chains.
We hide users' communication patterns by
ensuring every user is equally likely to be talking to any other user,
and hiding the origins of users' messages through mix-nets.
We then protect against active attacks efficiently using a novel
technique called aggregate hybrid shuffle.
Our evaluation on a network of 100 servers
demonstrates that \sys can support 2 million users
in 251 seconds,
which is more than $3.7\times$ faster than prior works with similar guarantees.

\bibliographystyle{plain}
\bibliography{refs}

\appendix
\section{Security of aggregate hybrid shuffle} \label{app:active_security}
The adversary's goal is to have an upstream server
successfully tamper with some messages
without getting detected by the honest server.
To model this, we consider the following security game between
three parties: the client, the adversary, and the verifier.
All parties are given the total number of users $\users$.
The client controls users in set $X_H \subset [\users]$ (this models the honest users),
and the adversary controls users in set $X_A = [\users] \setminus X_H$.
In addition, the adversary controls
servers $1, \ldots, h-1$,
and the verifier controls server $h$ (i.e., the honest server).
To simplify the presentation, we assume the adversary uses
the identity permutation for all servers,
but it is easy to adapt this proof to any arbitrary permutation.

\begin{enumerate}
  \item The adversary sends the client and the verifier
    the public keys $\blindkey_i$ and $\mixingkey_i$
    and $\innerkey_i$ for $i = 1, \ldots, h-1$.
    It also generates NIZKs to prove that it knows the values
    ${\privateblindkey_i = \log_{\blindkey_{i-1}}(\blindkey_i)}$
    and ${\privatemixingkey_i = \log_{\blindkey_{i-1}}(\mixingkey_i)}$
    for $i = 1, \ldots, h-1$,
    where $\blindkey_0 = g$.
    It sends the public keys and NIZKs to the verifier.
  \item The verifier verifies the NIZKs.
    The verifier then generates the key pairs
    $(\blindkey_h = \blindkey_{i-1}^{\privateblindkey_h}, \privateblindkey_h)$,
    ${(\mixingkey_h = \blindkey_{i-1}^{\privatemixingkey_h}, \privatemixingkey_h)}$,
    and $(\innerkey_h, \privateinnerkey_h)$,
    and sends the public keys to the client and the adversary.
  \item The client generates random $\{x_j\}_{j \in X_H}$,
    and ${\{\outerc^j = (X^j_1 = g^{x_j}, \outerc_1^j)\}_{j \in X_H}}$
    using the protocol described in \S\ref{sec:active_clients}.
    It also generates a NIZK that it knows corresponding to
    $x_j$ for each $j$, and sends both $\outerc^j$ and
    the NIZK to the adversary and the verifier.
  \item \label{step:adv_input}
    The adversary generates its input messages
    ${\{\outerc^j = (X^j_1, \outerc_1^j)\}_{j \in X_A}}$
    (not necessarily by following the protocol in \S\ref{sec:active_clients}).
    It generates a NIZK that shows it knows the discrete log of $X^j_1$
    and sends $\{\outerc^j\}_{j \in X_A}$ and the NIZKs
    to the client and the verifier.
  \item The verifier verifies all NIZKs.
  \item For $i = 1, \ldots, h-1$,
    the adversary sends the verifier $\{X^j_{i+1}\}_{j \in [\users]}$,
    and a NIZK that shows
    \[
      \left(\prod_{j=1}^\users X^j_i\right)^{\privateblindkey_i} = \prod_{j=1}^\users X^j_{i+1}
      \]
    by proving that
    \[
    \log_{\prod_{j=1}^\users X^j_i}\left(\prod_{j=1}^\users X^j_{i+1}\right)
    = \log_{\blindkey_{i-1}}(\blindkey_i) \ .
    \]
    It also sends the ciphertexts $\{{\outerc^j_h}'\}$ to the verifier.
  \item The verifier verifies all NIZKs,
    and checks that $\adec((X^j_h)^{\privatemixingkey_h}, {\outerc^j_h}') = (1, \cdot)$
    for all $j \in [\users]$.
\end{enumerate}
The game halts if the verifier fails to verify any NIZKs or
authenticated decryption ever fails (i.e., returns $(0, \cdot)$).
The adversary wins the game if the game does not halt
and it has successfully tampered with some messages.
In other words, the adversary wins if
\begin{enumerate}
  \item $\left(\prod_{j=1}^\users X^j_{i}\right)^{\privateblindkey_i} =
    \prod_{j=1}^\users X^j_{i+1}$ for all $i = 1, \ldots, h-1$,
  \item there exists $X_T \subset X_H$ such that
    for all $j \in X_T$,
    $(X^j_1)^{\prod_{i < h} \privateblindkey_i} \neq X^j_h$
    and $|X_T| > 0$,
  \item and $\adec((X^j_h)^{\privatemixingkey_h}, {\outerc^j_h}') = (1, \cdot)$
    for all $j \in [\users]$.
\end{enumerate}

We will now show that if the adversary can win this game,
then it can also break Diffie-Hellman.
Assume the adversary won the game.
Let $\privateblindkey_A = \prod_{i < h} \privateblindkey_i$ be the product
of the private blinding key of the adversary.
If the adversary won, then the first condition implies that
${(\prod_{j=1}^\users X^j_1)^{\privateblindkey_A} = \prod_{j=1}^\users X^j_h}$.
Now, consider three boolean predicates for each $j \in [\users]$:
$\outerc^j_h \stackrel{?}{=} {\outerc^j_h}'$,
$(X^j_1)^{\privateblindkey_A} \stackrel{?}{=} X^j_h$,
and $\know((X^j_h)^{\privatemixingkey_h})$,
where $\know(x) = 1$ if the adversary knows (or can compute) $x$,
and 0 otherwise. There are eight possible combinations of the predicates,
and we consider each combination for $j \in X_H$.
We indicate which combinations are possible for the adversary to satisfy,
given that all authenticated decryptions were successful.
\begin{enumerate}
  \item \textbf{$\outerc^j_h \neq {\outerc^j_h}'$,
    $(X^j_1)^{\privateblindkey_A} \neq X^j_h$,
    $\know((X^j_h)^{\privatemixingkey_h}) = 0$: IMPOSSIBLE.}
    Since the adversary does not know the key used to decrypt
    (i.e., $(X^j_h)^{\privatemixingkey_h}$),
    it cannot generate a valid ciphertext.
  \item \textbf{$\outerc^j_h \neq {\outerc^j_h}'$,
    $(X^j_1)^{\privateblindkey_A} \neq X^j_h$,
    $\know((X^j_h)^{\privatemixingkey_h}) = 1$: POSSIBLE.}
    Since the adversary knows the key used to decrypt
    it could generate a valid ciphertext.
  \item \textbf{$\outerc^j_h \neq {\outerc^j_h}'$,
    $(X^j_1)^{\privateblindkey_A} = X^j_h$,
    $\know((X^j_h)^{\privatemixingkey_h}) = 0$: IMPOSSIBLE.}
    Same argument as case 1.
  \item \textbf{$\outerc^j_h \neq {\outerc^j_h}'$,
    $(X^j_1)^{\privateblindkey_A} = X^j_h$,
    $\know((X^j_h)^{\privatemixingkey_h}) = 1$: IMPOSSIBLE.}
    If possible, then the adversary can break the Diffie-Hellman assumption.
    Namely, given only $X^j_1 = g^{x_j}$, $\privateblindkey_A$,
    and $(g^{\privateblindkey_A})^{\privatemixingkey_h}$ for random $x_j$ and $\privatemixingkey_h$ and an independently generated $\privateblindkey_A$,
    it can compute $(X^j_h)^{\privatemixingkey_h} = g^{x_j \cdot \privateblindkey_A \cdot \privatemixingkey_h}$.
    If this were possible, then given $g^a$ and $g^b$ for random $a$ and $b$,
    the adversary could generate an $\privateblindkey_A$,
    compute $(g^b)^{\privateblindkey_A}$,
    and compute $g^{a \cdot b \cdot \privateblindkey_A}$.
    It could then break the Diffie-Hellman assumption and compute $g^{ab}$
    by raising $g^{a \cdot b \cdot \privateblindkey_A}$ to $\privateblindkey_A^{-1}$.
  \item \textbf{$\outerc^j_h = {\outerc^j_h}'$,
    $(X^j_1)^{\privateblindkey_A} \neq X^j_h$,
    $\know((X^j_h)^{\privatemixingkey_h}) = 0$: IMPOSSIBLE.}
    If possible, then $\outerc^j_h$ authenticates under two different keys
    $(X^j_1)^{\privateblindkey_A \cdot \privatemixingkey_h}$
    and $(X^j_h)^{\privatemixingkey_h}$.
    However, the probability of this is negligible (\S\ref{sec:model}).
  \item \textbf{$\outerc^j_h = {\outerc^j_h}'$,
    $(X^j_1)^{\privateblindkey_A} \neq X^j_h$,
    $\know((X^j_h)^{\privatemixingkey_h}) = 1$: IMPOSSIBLE.}
    Same argument as case 5.
  \item \textbf{$\outerc^j_h = {\outerc^j_h}'$,
    $(X^j_1)^{\privateblindkey_A} = X^j_h$,
    $\know((X^j_h)^{\privatemixingkey_h}) = 0$: POSSIBLE.}
    This corresponds to untampered messages.
  \item \textbf{$\outerc^j_h = {\outerc^j_h}'$,
    $(X^j_1)^{\privateblindkey_A} = X^j_h$,
    $\know((X^j_h)^{\privatemixingkey_h}) = 1$: IMPOSSIBLE.}
    Same argument as case 4.
\end{enumerate}
Thus, there are two possible combinations of predicates:
($\outerc^j_h \neq {\outerc^j_h}'$,
$(X^j_1)^{\privateblindkey_A} \neq X^j_h$,
$\know((X^j_h)^{\privatemixingkey_h}) = 1$)
and
($\outerc^j_h = {\outerc^j_h}'$,
$(X^j_1)^{\privateblindkey_A} = X^j_h$,
$\know((X^j_h)^{\privatemixingkey_h}) = 0$).
The first combination corresponds exactly to $X_T$ (tampered messages),
and the second corresponds exactly to $X_H \setminus X_T$ (untampered messages).

Similarly, consider $j \in X_A$.
The adversary generates the ciphertexts $\{{\outerc^j_h}'\}$
for the verifier. Thus, the adversary must know $(X^j_h)^{\privatemixingkey_h}$,
the key used to authenticate the ciphertext, for $j \in X_A$.

Now, we consider the product of the users' Diffie-Hellman keys.
Because all NIZKs have to be verified, we have that
\[
\left(\prod_{j=1}^\users X^j_1\right)^{\privateblindkey_A} =
\left(\prod_{j=1}^\users X^j_h\right) \ .
\]
Consider $X_U = X_H \setminus X_T$, i.e., the set of messages that did not change.
Then, we can divide both sides by the values associated with $X_U$
since $(X^j_1)^{\privateblindkey_A} = X^j_h$ for $j \in X_U$:
\[
  \left(\prod_{j \in X_T \cup X_A} X^j_1\right)^{\privateblindkey_A} =
  \left(\prod_{j \in X_T \cup X_A} X^j_h\right) \ ,
\]
since $X_T \cup X_A = [\users] \setminus X_U$.
We can rewrite this as
\begin{equation} \label{eq:active_eq}
  \left(\prod_{j \in X_T} X^j_1\right)^{\privateblindkey_A} =
  \left(\prod_{j \in X_T \cup X_A} X^j_h\right) /
  \left(\prod_{j \in X_A} X^j_1\right)^{\privateblindkey_A} \ .
\end{equation}
Based on our analysis of the possible predicates for $X_T$ and $X_A$,
the adversary must know $(X^j_h)^{\privatemixingkey_h}$ for $j \in X_T \cup X_A$.
Moreover, the adversary knows $\log_g(X^j_1)$ for $j \in X_A$
(it was required to prove the knowledge in step~\ref{step:adv_input} of the game).
Thus, the adversary can compute
$((X^j_1)^{\privateblindkey_A})^{\privatemixingkey_h}$
by computing $((g^{\privateblindkey_A})^{\privatemixingkey_h})^{\log_g(X^j_1)}$
for $j \in X_A$
(it knows $\mixingkey_h = (g^{\privateblindkey_A})^{\privatemixingkey_h}$).
As a result, it can compute
\[
  \left(\prod_{j \in X_T \cup X_A} \left(X^j_h\right)^{\privatemixingkey_h}\right) /
  \prod_{j \in X_A} \left(\left(X^j_1\right)^{\privateblindkey_A}\right)^{\privatemixingkey_h} \ .
\]
This, however, is
\begin{align*}
  & \left(\prod_{j \in X_T \cup X_A} \left(X^j_h\right)^{\privatemixingkey_h}\right) /
  \prod_{j \in X_A} \left(\left(X^j_1\right)^{\privateblindkey_A}\right)^{\privatemixingkey_h} \\
  & = \left(\left(\prod_{j \in X_T \cup X_A} X^j_h\right) /
  \left(\prod_{j \in X_A} X^j_1\right)^{\privateblindkey_A} \right)^{\privatemixingkey_h} \\
  & = \left(\left(\prod_{j \in X_T} X^j_1\right)^{\privateblindkey_A}\right)^{\privatemixingkey_h} \ ,
\end{align*}
where the last step uses the equality from equation~\ref{eq:active_eq}.
This means that given $\{X^j_1 = g^{x_j}\}$, $\privateblindkey_A$,
and $(g^{\privateblindkey_A})^{\privatemixingkey_h}$
for random $\{x_j\}$ and $\privatemixingkey_h$,
and $\privateblindkey_A$ that was generated independently of $\privatemixingkey_h$,
the adversary was able to compute
$g^{\privateblindkey_A \cdot \privatemixingkey_h \cdot \sum_{j \in X_T} x_j}$.

However, such an adversary can also break Diffie-Hellman.
To see how, consider the following adversary $\adv_{\kex}$ that tries
to break Diffie-Hellman.
$\adv_{\kex}$ is given $g^a$ and $g^b$ for random $a$ and $b$,
and is asked to compute $g^{ab}$.
To compute this, $\adv_{\kex}$ plays the above game setting
$g^{x_1} = g^a$ and $\mixingkey_h = (g^{\privateblindkey_A})^b = (g^b)^{\privateblindkey_A}$
for $\privateblindkey_A$ of its choosing, and then simulates the client
by generating many random $x_j$ for $j \in X_H \setminus \{1\}$.
At the end of the game, the adversary will have
$g^{\privateblindkey_A \cdot b \cdot (a + \sum_{j \in X_T, j \neq 1} x_j)}$.
Since the adversary knows $\privateblindkey_A$ and $x_j$ for $j \neq 1, j \in X_T$,
it can compute ${g^{ab} = (g^{\privateblindkey_A \cdot b \cdot (a + \sum_{j \in X_T, j \neq 1} x_j)})^{\privateblindkey_A^{-1}} / (g^b)^{\sum_{j \in X_T, j \neq 1} x_j}}$.
Thus, it can break Diffie-Hellman.
Therefore, the adversary cannot win the above game if Diffie-Hellman is hard,
meaning that it could satisfy at most two out of the three conditions
to win the game.
In turn, this implies that the honest server will always catch
an upstream malicious server misbehaving.

\section{Security game and proof sketches} \label{sec:security}
We define the security of our system using the following
security game played between a challenger and an adversary.
Both the challenger and the adversary are given
the set of users $[\users] = \{1,2,\ldots,\users\}$,
the set of servers $[\servers]$,
the fraction $f$ of servers the adversary can compromise,
and the number of chains $\chains$.
\begin{enumerate}
  \item \label{game:sel_servers}
    The adversary selects the set of malicious servers
    ${A_s \subset [\servers]}$ such that $|A_s| \leq f \cdot \servers$,
    and the set of malicious users
    $A_c \subset [\users]$ such that $|A_c| \leq \users - 2$.
    The adversary sends $A_s$ and $A_c$ to the challenger.
    Let ${H_s = [\users] \setminus A_s}$ and
    ${H_c = [\servers] \setminus A_c}$ denote the set of honest
    servers and users.

  \item \label{game:create_chains}
    The challenger computes the size of each chain $\chain$
    as a function of $f$ and $\chains$, as described in \S\ref{sec:mixnet}.
    Then, it creates $\chains$ mix chains by repeatedly sampling $\chain$
    servers per group at random.
    The challenger sends the chain configurations to the adversary.

  \item \label{game:gen_keys}
    The adversary and the challenger generate
    blinding keys, mixing keys, and inner keys
    as described in \S\ref{sec:active_keys} and Appendix~\ref{app:active_security}.

  \item The adversary picks some honest users $H_t \subset H_c$
    such that $|H_t| \geq 2$.
    It generates sets of chains $\{C_x\}$ for $x \in H_t$
    such that $C_x \cap C_y \neq \emptyset$ for all $x,y \in H_t$.
    For $\prfchain \in [\chains]$,
    let ${U_\prfchain = \{x \in H_t : \prfchain \in C_x\}}$.
    For each chain $\prfchain$,
    it also generates the potential messages
    $\{\msg_{xy}^\prfchain\}$ for $x,y \in U_\prfchain$
    where $\msg_{xy}^\prfchain$ is the message that may be
    sent from user $x$ to user $y$ in chain $\prfchain$.
    The adversary sends $H_t$, $\{\{\msg_{xy}^\prfchain\}\}$,
    and $\{C_x\}$ to the challenger.

  \item \label{game:submit_user_msgs}
    The challenger first verifies that every pair of $C_x$ and $C_y$
    intersects at least once. If this is not the case, the game halts.
    The challenger then performs the following for each chain $\prfchain \in [\chains]$.
    First, it creates conversation pairs for each chain $\prfchain$
    $\{(X_i,Y_i)_\prfchain\} \subset U_\prfchain \times U_\prfchain$ at random such that
    every $x \in U_\prfchain$ appears in exactly one of the pairs.
    In other words, every user has a unique conversation partner per chain.
    (If $X_i = Y_i$, then that user is talking with herself.)
    For each ${(x,y) \in \{(X_i,Y_i)_\prfchain\}}$,
    the challenger onion-encrypts the messages
    $\msg_{xy}^\prfchain$ from $x$ to $y$ and
    $\msg_{yx}^\prfchain$ from $y$ to $x$
    with the keys of servers in chain $\prfchain$.
    Then, it uses the protocol described in \S\ref{sec:active_clients}
    to submit the ciphertexts and the necessary NIZKs.

  \item The adversary generates inputs to the chains for the users in $A_c$,
    and sends them to the chains.

  \item \label{game:shuffle}
    The challenger and the adversary take turns processing the messages
    in each chain. Within a chain, they perform the following
    for $i = 1, \ldots, \chain$:
    \begin{enumerate}
      \item If server $i \in H_s$, the challenger performs protocol
        described in \S\ref{sec:active_mixing}
        to shuffle and decrypt the messages, and also generate an AHS proof.
        The challenger then sends the proof to the adversary,
        and the resulting messages to the owner of server $i+1$.
      \item If server $i \in A_s$, the adversary generates some messages
        along with an AHS proof.
        Then, sends the AHS proof to the challenger,
        and sends the messages to the owner of server $i+1$.
    \end{enumerate}
    The challenger verifies all AHS proofs.

  \item \label{game:inner_decrypt}
    The challenger and the adversary decrypt
    the final result of the shuffle (i.e., the inner ciphertexts).

  \item \label{game:random_b}
    The challenger samples a random bit $b \leftarrow {0,1}$.
    If $b=0$, then send the adversary $\{(X_i,Y_i)_\prfchain\}$
    for ${\prfchain \in [\chains]}$.
    If $b=1$, then sample random conversation pairs
    ${\{(X_i',Y_i')_\prfchain\} \subset U_\prfchain \times U_\prfchain}$
    for each chain
    with the same constraint as in step~\ref{game:submit_user_msgs},
    and send the adversary the newly sampled pairs.

  \item The adversary makes a guess $b'$ for $b$.
\end{enumerate}
The adversary wins the game if the game does not
come to a halt before the last step and $b'=b$.
The adversary need not follow the protocol described in this paper.
The advantage of the adversary in this game is
$|\Pr[b' = b] - \frac{1}{2}|$.
We say that the system provides metadata private communication
if the advantage is negligible in the implicit security parameter.
Note that this game models a stronger version of \sys,
which allows users to communicate with multiple users
on different chains (i.e., what is described in \S\ref{sec:discuss}),
rather than only one user.
We could change the game slightly to force the challenger
to send loopback messages in step~\ref{game:submit_user_msgs}
to model having just one conversation.

\paragraph{Proof sketches.}
First, we argue that the adversary needs to tamper
with messages prior to the last honest server shuffling,
as stated in \S\ref{sec:active}. To see why,
consider an adversary that only tampers with the messages after
the last honest server. The adversary can learn
the recipients of all messages, but not the senders.
As a result, the adversary does not learn anything about whether two users
$x, y \in U_\prfchain$ received messages because there exists
a conversation pair $(x,y)_\prfchain$,
or because there were two conversation pairs
$(x,x)_\prfchain$ and $(y,y)_\prfchain$.
This means that any set of conversation pairs is equally likely
to be sampled by the challenger from the adversary's view,
meaning that the adversary does not gain any advantage.
Thus, we consider an adversary who tampers with messages
prior to the honest server processing the messages.

In this scenario, the adversary in step~\ref{game:shuffle}
must follow the protocol (e.g., no tampering with the messages),
as analyzed in Appendix~\ref{app:active_security}.
Given this restriction, we now argue that the adversary does not learn anything
after playing the security game
by describing how a simulator of an adversary could simulate the whole game
with only the public inputs and the private values of the adversary.

The simulator can simulate step~\ref{game:gen_keys} by generating
random public keys.
It can simulate step~\ref{game:submit_user_msgs}
by generating random values in place of the ciphertexts that
encrypt the users' messages $\{\{\msg_{xy}^\prfchain\}\}$,
since the ciphertexts are indistinguishable from random.
It then randomly matches a user in $U_\prfchain$ to one of the
generated random values for each chain $\prfchain$, and
sets the destination of each message as the matched user.
It onion-encrypts the final message using the randomly generated public keys
and the adversary's public keys.
In step~\ref{game:shuffle}, the adversary simulates the challenger
by randomly permuting the messages,
and removing a layer of the encryption from the messages.
(It can remove a layer of encryption since
it knows all layers of onion-encryption.)
Finally, it could simulate the challenger's last challenge
by picking sets of randomly generated conversation pairs,
subject to the constraints in step~\ref{game:submit_user_msgs}.
The distribution of the messages generated and exchanged in the security game
and in the simulator are indistinguishable for a computationally limited adversary.

\end{document}